\documentclass[preprintnumbers,amssymb]{revtex4}

\usepackage{graphicx,epsfig}% Include figure files
\usepackage{bm}

\newcommand{\be}{\begin{equation}}
\newcommand{\ee}{\end{equation}}
\newcommand{\bea}{\begin{eqnarray}}
\newcommand{\eea}{\end{eqnarray}}
\newcommand{\ba}{\begin{array}}
\newcommand{\ea}{\end{array}}
\newcommand{\beqn}{\begin{eqnarray*}}
\newcommand{\eeqn}{\end{eqnarray*}}

\def\nn{\nonumber}

\def\ii{{\rm i}}

\begin{document}

\nopagebreak

\title{Black hole quasinormal modes: hints of quantum gravity?}
\author{Emanuele Berti}

\affiliation{E-mail: berti@wugrav.wustl.edu \\ 
\it {McDonnell Center for the Space Sciences, 
Department of Physics, Washington University, 
St. Louis, Missouri 63130, USA} }

\begin{abstract}
\noindent

This is a short review of the quasinormal mode spectrum of
Schwarzschild, Reissner-Nordstr\"om and Kerr black holes, summarizing
results obtained in \cite{BK,BCKO,BCY}. The summary includes
previously unpublished calculations of i) the eigenvalues of
spin-weighted spheroidal harmonics, and ii) quasinormal frequencies of
extremal Reissner-Nordstr\"om black holes.
%We also compute quasinormal frequencies of charged {\it and} rotating
%black holes neglecting the electromagnetic perturbations. For these
%``frozen'' Kerr-Newman black holes the perturbation equations become
%separable. We consider the combined effect of angular momentum and
%charge on moderately damped quasinormal frequencies and discuss the
%limits of our approximation.

\end{abstract}
\maketitle

% {\it ``All right,'' said the cat; and this time it vanished quite
% slowly, beginning with the end of the tail, and ending with the grin,
% which remained some time after the rest of it had gone.

% ``Well! I've often seen a cat without a grin,'' thought Alice; ``but a
% grin without a cat! It's the most curious thing I ever saw in all my
% life!''} (Lewis Carroll)
% \vskip 12pt

% {\it By reason of its faster and faster infall [the surface of an
% imploding star] moves away from the distant observer more and more
% rapidly. The light is shifted to the red. It becomes dimmer
% millisecond by millisecond, and in less than a second is too dim to
% see... [The star,] like the Cheshire cat, fades from view. One leaves
% behind only its grin, the other only its gravitational attraction.}
% (John A. Wheeler)
% \vskip 12pt

\section{Introduction}
\label{sec:intro}

Black holes are one of the most extreme predictions of Einstein's
general relativity. They challenge common sense to such an extent that
Einstein himself was reluctant to accept their reality. Despite the
initial resistance, nowadays the existence of black holes is widely
accepted by the physics community. Observationally, the astronomical
evidence for black holes is compelling. Theoretically, a key step to
accept the idea that black holes are not just a mathematical solution
of the Einstein equations {\it in vacuo} was the proof of the (linear)
stability of the Schwarzschild metric. This proof is based on black
hole perturbation theory, as developed in two classic papers by
Regge-Wheeler and Zerilli \cite{RW, Zer}. Their formalism gets rid of
the angular dependence of the perturbation variables through a
tensorial generalization of the spherical harmonics, thus reducing the
solution of the perturbed Einstein equations to the study of a
Schr\"odinger-like wave equation with some potential $V(r)$. Stability
can then be established by methods that are familiar from quantum
mechanics: perturbations due to an external (eg., electromagnetic or
gravitational) field are considered as waves scattering off the
corresponding potential \cite{FHM}. It is worth stressing that the
causally disconnected region {\it inside} the event horizon plays no
role whatsoever in this analysis.

The Regge-Wheeler/Zerilli formalism was later extended to encompass
charged (Reissner-Nordstr\"om) or rotating (Kerr) black hole
solutions. For Reissner-Nordstr\"om (RN) black holes we have {\it two}
coupled Schr\"odinger equations, describing the mutual interaction of
gravitational and electromagnetic perturbations. For Kerr black holes
the problem is highly non-trivial, since the background geometry is
not spherical and tensor spherical harmonics cannot be used any
more. Nonetheless, Teukolsky was able to separate the angular and
radial dependence of the perturbations using a certain class of
special functions (the {\it spin-weighted spheroidal harmonics}; see
section \ref{sec:kerrae}) - a truly remarkable achievement
\cite{teukolsky}. A complete discussion of black hole perturbation
theory can be found in the monumental monograph by Chandrasekhar
\cite{MTB}. The extension to charged {\it and} rotating (Kerr-Newman)
black holes turned out to be a formidable task: the numerous attempts
to separate the corresponding perturbation equations have all failed.

Once we know that a black hole solution {\it is} indeed stable, the
next step is to understand``how much stable'' it is. In other words,
on which timescale does a black hole radiate away its ``hair'' after
formation? The answer to this question is provided by the black hole's
{\it quasinormal mode spectrum}. After an initial phase depending on
the details of the collapse, the black hole starts vibrating into
``quasinormal'' (exponentially decaying) oscillation modes whose
frequencies and decay times depend only on the intrinsic features of
the black hole itself, being insensitive to the details of the
collapse. 

% Quasinormal modes have been compared to the grin of the Cheshire cat
% \cite{Nils}: they are all that's left of the collapsing matter once it
% crosses the black hole's event horizon and becomes causally
% disconnected from the external world. A study of the quasinormal
% spectrum reserves many surprises. If the quasinormal spectrum of black
% holes is their ``grin'', it is indeed ``a curious thing''.

We have seen that for Schwarzschild, RN and Kerr black holes the
perturbation problem can be reduced to the study of certain
characteristic wave equations. The background spacetime curvature
determines the shape of the potential in each wave
equation. Quasinormal modes are defined as solutions that satisfy
``natural'' boundary conditions for classical waves: they are purely
outgoing at spatial infinity and purely ingoing at the black hole
horizon. In the Fourier domain, this is an eigenvalue problem for the
wave's (complex) frequency $\omega=\omega_R+\ii \omega_I$. The two
boundary conditions are satisfied by a discrete set of quasinormal
frequencies, $\omega=\left\{\omega_n\right\}$, where the integer
$n=0,1,\dots$ labels the modes, which are normally sorted by
increasing values of $|\omega_I|$. Quasinormal frequencies are complex
({\it quasi-}normal) because outgoing waves at infinity radiate energy
(the system is not conservative). The damping time of each oscillation
is of course given by $\tau=1/|\omega_I|$: modes having large values
of $|\omega_I|$ damp faster. The quasinormal mode spectrum, unlike
ordinary normal mode expansions, is not complete: in general, we
cannot express a given perturbation as a superposition of quasinormal
modes (see section 4.4 of reference \cite{NR} for a
discussion). Completeness would be desirable from a mathematical point
of view, but in any case quasinormal mode expansions are useful (and
routinely used) to describe the final stages of formation of a black
hole after collapse. For example, it has been shown that the numerical
waveforms obtained from general relativistic simulations of distorted
black holes are very well fitted using the fundamental quasinormal
mode, with the possible addition of a few overtones \cite{BS}.

The rest of this short review is devoted to a detailed presentation of
the quasinormal mode spectrum of Schwarzschild, RN and Kerr black
holes. The quasinormal spectrum has a complicated structure: as it
happens, black hole perturbation theory leads us once again ``into a
realm of the rococo: splendorous, joyful, and immensely ornate''
\cite{MTB}. Such an ornate spectrum can at times be hard to compute,
even by the most sophisticated numerical methods. 

In particular, for a long time the high-damping limit of the spectrum
has been a no-go zone. From a physical point of view, it is clear that
slowly-damped quasinormal modes should dominate the response of a
black hole to any type of perturbation: this obvious consideration
initially reduced the impetus to explore the large-$|\omega_I|$ (high
damping) regime. From a technical perspective, most numerical methods
to impose the quasinormal mode boundary conditions are based on
discriminating outgoing waves from ingoing waves. The exponentially
decaying solution is quickly swamped into numerical noise, and this
effect is larger for larger values of the damping. As a consequence,
all numerical methods are doomed to failure at some critical value of
$|\omega_I|$.

More recently, the high-damping region of the quasinormal mode
spectrum gained more attention because of a conjectured connection
with quantum gravity. York \cite{Y} first attempted to relate the
(purely classical) quasinormal mode spectrum with quantum properties
of black holes, and more specifically Hawking radiation. In 1998 Hod
\cite{hod} noticed that, as $|\omega_I|\to \infty$, the numerically
computed oscillation frequencies of Schwarzschild black holes tend to
a constant asymptotic value $\omega_R^*=T_H \ln 3$, where $T_H$ is the
Hawking temperature of the black hole. In the same limit, the
(constant) spacing between modes is given by $2\pi T_H$. Assuming that
the minimum energy emitted or absorbed by a black hole is $\Delta
M=\hbar \omega_R^*$ and using the Bekenstein-Hawking area-entropy
relationship, Hod found an equally spaced black hole area spectrum (in
agreement with some heuristic arguments by Bekenstein \cite{bek}). By
virtue of the ``$\ln 3$'' factor, the resulting spectrum is also
consistent with a statistical-mechanical interpretation of black hole
entropy. The idea was revived more recently by Dreyer \cite{dreyer},
who used a similar argument to fix the Barbero-Immirzi parameter
\cite{dreyer} - an undetermined constant of proportionality entering
black hole entropy calculations in Loop Quantum Gravity.

The results we show here do {\it not} seem to support an extension of
Hod's original proposal to charged and rotating black holes. In
addition, the original Loop Quantum Gravity calculation of the black
hole area spectrum was recently re-examined and shown to be in error
by Domagala and Lewandowski \cite{DL}. These developments could be
used to argue against the relevance of quasinormal modes for Loop
Quantum Gravity, or any other theory of quantum gravity for that
matter. Some hints to clarify -- or kill once and for all -- the
connection between quasinormal modes and area quantization might come
from analogue black holes \cite{BCL}, but in that case the
thermodynamical interpretation of horizon area as entropy is not a
trivial issue.

Whether or not quasinormal frequencies have something to say about
quantum gravity, the recent conjectures stimulated a lot of work to
clarify the structure of the high-damping region of the quasinormal
mode spectrum. The following sections summarize some of these recent
developments. Our aim is {\it not} to give a complete overview of
recent work in this field: that would be an ambitious goal. For
example, we do not discuss recent progress on the asymptotic
quasinormal spectrum of black holes in non-asymptotically flat
spacetimes or higher dimensions (with a few exceptions where
appropriate). We only present a personal (and therefore biased)
account of some open problems, concentrating on the Kerr-Newman family
of black hole solutions. Before turning to a description of the
quasinormal mode spectrum, it is useful to mention in passing
different numerical methods that have been used in the past for
quasinormal mode calculations.

\section{Numerical methods}

The summary we present in this section is only meant to highlight pros
and cons of some popular numerical methods to compute quasinormal
modes, and ultimately to justify our use of the continued fraction
technique in the present context. The list is by no means
exhaustive. For a more complete discussion of the merits and drawbacks
of each method we refer the reader to section 6.1 of \cite{KS}.

\begin{itemize}

\item[1)] {\it Time-domain evolutions} of the Regge-Wheeler and
Zerilli equations were first performed by Vishveshwara \cite{vish} and
Press \cite{press}: this is actually how quasinormal modes were first
discovered. However, finding highly damped modes in this way requires
long and stable evolutions. This method is not used any more for
quasinormal mode calculations, with some notable exceptions in cases
where a Fourier-domain treatment is cumbersome or impossible. Examples
include i) the simulation of complicated distributions of matter in a
fixed (black hole or stellar) background \cite{nagar}, and ii) time
evolutions of the Kerr perturbation equations (see \cite{lousto} and
references therein). The latter are particularly important for the
simulation of extreme mass ratio inspirals, one of the target sources
for the space-based gravitational wave interferometer LISA. No
generalization of the Carter constant is known for off-equatorial,
non-circular particle orbits in the Kerr metric, and time evolutions
provide a valid alternative to calculations of radiation reaction
effects using self-force methods.

\item[2)] {\it Direct integrations of the wave equation in the
frequency domain} were pioneered by Chandrasekhar and Detweiler
\cite{cd}. The simplest version of this approach is numerically
unstable: the exponentially vanishing solution is very hard to extract
from numerical noise in the exponentially growing solution, a problem
that plagues especially highly damped modes. Several improvements have
been attempted. A notable one was introduced by Nollert and Schmidt,
who used a sophisticated Laplace transform definition of the
quasinormal frequencies \cite{ns}. In this framework, the state of the
art is probably the complex integration technique developed by
Andersson and coworkers \cite{nilsci}. In any case, direct
integrations are quite time consuming. Phase-integral based methods
(see below) sometimes involve numerical integrations; they can be
competitive with continued fraction methods at moderate dampings, but
ultimately fail at very high values of $|\omega_I|$.

\item[3)] {\it Inverse potential methods} have been introduced by
Mashoon and collaborators \cite{fm}. The idea is to approximate the
exact potential by some simpler potential (eg. a potential of the
P\"oschl-Teller form) whose spectrum can be found analytically. These
techniques can be a valuable tool in certain situations \cite{CL}, but
in general results obtained from approximating potentials should not
be expected to be accurate. For example, Nollert showed that
approximating the Regge-Wheeler potential by a series of square wells
of decreasing width does not yield the expected quasinormal
frequencies as the width of the square wells tends to zero. Analytic
approximations can still be useful in certain limits, such as the
eikonal approximation (large-$l$ limit) or the high damping limit
\cite{M,MN,N2}.

\item[4)] {\it WKB methods} were originally suggested by Schutz and
Will \cite{sw} and are based on elementary quantum mechanical
arguments. The analogy between the wave equation describing black hole
perturbations and the Schr\"odinger equation can be exploited to use
standard WKB expansions. At lowest order, quasinormal frequencies
are given by the Bohr-Sommerfeld rule:
\be
\int_{r_A}^{r_B}\left[\omega^2-V(r)\right]^{1/2}dr=(n+1/2)\pi\,,
\ee
where $r_A$ and $r_B$ are the two roots (turning points) of
$\omega^2-V(r)=0$.  The WKB expansion can be pushed to higher orders
\cite{iw,i}, allowing a simple and reasonably accurate determination
of the quasinormal frequencies and a trasparent physical
interpretation of the results. The method can be generalized to deal
with RN \cite{ks} and even Kerr black holes \cite{si}. In the latter
case, unfortunately, the WKB method is not very simple to apply (the
Kerr potential in the standard Teukolsky equation is complex in the
first place, but it can be made real transforming the equation to a
different form) and does not yield very accurate results. The WKB
approximation gets better for large values of $l$, allowing an
analytic determination of quasinormal frequencies in the large-$l$
limit, but it was found to fail badly at high values of the damping
\cite{guinn}.

\item[5)] {\it Phase-integral methods} are a substantial improvement
over standard WKB techniques \cite{phase,A2,ga}. Their major drawback
is related to the fact that they require integrations in the complex
plane. The integration contour must be found case by case, and the
problem can be quite tricky, depending on the singularity structure of
the potential. A variant of the phase-integral method has been used
to compute analytically the Schwarzschild and RN quasinormal frequency
in the limit $|\omega_I|\to \infty$ \cite{AH}. A similar analytical
treatment for the Kerr case is still missing.

\item[6)] {\it Continued fractions} were first used by Leaver \cite{L}
exploiting once again an analogy with quantum mechanics. Leaver's
approach is based on a classical 1934 paper by Jaff\'e on the
electronic spectra of the H$^+$ ion \cite{jaffe}. The key steps are:
a) recognize that the wave equations can be seen as special cases of
generalized spheroidal wave equations; b) write appropriate series
expansions for the solutions; c) replace the series expansions in the
differential equations and derive recursion relations for the
expansion coefficients; d) use the recursion relations to study the
convergence properties of these series and impose quasinormal mode
boundary conditions. The determination of quasinormal frequencies
boils down to a numerical solution of continued fraction relations
involving the mode frequency and the black hole parameters. The
evaluation of continued fractions only involves elementary algebraic
operations and the convergence of the method is excellent, even at
high damping. In the Schwarzschild case, in particular, the method can
be tweaked to allow the determination of modes of order up to $\sim
100,000$ \cite{N}.  Gauss-Jordan elimination techniques can be used to
encompass the RN \cite{L2} and even extreme RN \cite{OMOI} cases.

\end{itemize}

Other approaches may be preferred in specific situations, but Leaver's
continued fraction technique is the best workhorse method to compute
highly damped modes in general situations. It is very reliable,
largely independent of the form of the potential (that is only used to
find the recursion coefficients once and for all) and fast, bypassing
numerical integrations of the wave equations.

All the results we present in the following have been obtained using
different variants of Leaver's method. We adopt geometrical units and
consistently use Leaver's conventions \cite{L}; in particular, we set
$2M=1$. This means that extremal Kerr and RN black holes correspond,
respectively, to $a=1/2$ and $Q=1/2$.

%%%%%%%%%%%%%%%%%%%%%%%%%%%%%%%%%%%%%%%%%%%%%%%%%%%%%%%%%%%%%%%%%%%%%%%%%%%%%%%
%%%%%%%%%%%%%%%%%%%%%%%%%%%%%%%%%%%%%%%%%%%%%%%%%%%%%%%%%%%%%%%%%%%%%%%%%%%%%%%

\section{Schwarzschild black holes}
\label{sec:schw}

\subsection{Computational method}

Let us begin from the simplest case of spherically symmetric,
uncharged, non-rotating Schwarzschild black holes. A detailed
derivation of the perturbation equations can be found in \cite{MTB}
(for more details on the computational procedure see \cite{L,N}). In
brief: the angular dependence of the metric perturbations can be
separated using tensorial spherical harmonics \cite{Zer}. Depending on
their behavior under parity, the perturbation variables are classified
as {\it polar} (even) or {\it axial} (odd). We can get rid of the time
dependence using a standard Fourier decomposition for each
perturbation variable $\psi$:
\be
\psi(t,r)=\frac{1}{2\pi}\int_{-\infty}^{+\infty}
e^{-\ii \omega t}\psi(\omega,r) d\omega\,.
\ee
The resulting differential equations can be manipulated to yield two
wave equations, one for the polar perturbations (that we shall denote
by a superscript plus) and one for the axial perturbations
(superscript minus):
\be\label{wave}
\left({d^2\over dr_*^2}+\omega^2\right)Z^{\pm}=V^{\pm} Z^{\pm}\,,
\ee
where we introduced a tortoise coordinate $r_*$ defined in the usual
way by the relation
\be\label{tortoise}
{d r\over d r_*}={\Delta\over r^2}\,,
\ee
and $\Delta=r(r-1)$. 

Polar perturbations are related to axial perturbations by a
differential transformation discovered by Chandrasekhar \cite{MTB}. As
anticipated in the introduction, quasinormal modes are solutions of
equation (\ref{wave}) that are purely outgoing at spatial infinity
($r\to \infty$) and purely ingoing at the black hole horizon ($r\to
1$). These boundary conditions are only satisfied by a discrete set of
complex frequencies (the quasinormal frequencies). For the
Schwarzschild solution the two potentials $V^{\pm}$ are quite
different, yet the quasinormal modes for polar and axial perturbations
are the same \cite{MTB}. This can be seen as a consequence of the two
potentials being related by a supersymmetry transformation
\cite{SUSY}. Since polar and axial perturbations are isospectral and
$V^-$ has an analytic expression which is simpler to handle, we can
concentrate on the axial equation. Written in terms of the
``standard'' radial variable $r$, this equation reads:
\be\label{sax}
r(r-1)\frac{d^2\psi_l}{dr^2}+\frac{d\psi_l}{dr}-
\left[l(l+1)-\frac{s^2-1}{r}-\frac{\omega^2 r^3}{r-1}\right]\psi_l=0\,,
\ee
where $s$ is the spin weight of the perturbing field ($s=0,~-1,~-2$
for scalar, electromagnetic and gravitational perturbations,
respectively) and $l$ is the angular index of the perturbation. Notice
that perturbations of a Schwarzschild background are independent of
the azimuthal quantum number $m$, because the background spacetime has
spherical symmetry; the same is true in the RN case, but not for Kerr
black holes. Equation (\ref{sax}) can be solved using a series
expansion of the form:
\be\label{sser}
\psi_l=(r-1)^{-\ii \omega}r^{2\ii \omega}e^{\ii \omega (r-1)}
\sum_{j=0}^{\infty} a_j \left(\frac{r-1}{r}\right)^j\,.
\ee
where the prefactor is chosen to incorporate the quasinormal boundary
conditions at the two boundaries. Substituting the series expansion
(\ref{sser}) in (\ref{sax}) we get a three term recursion relation for
the expansion coefficients $a_j$:
\bea
&&\alpha_0 a_1+\beta_0 a_0=0,\\
&&\alpha_j a_{j+1}+\beta_j a_j+\gamma_j a_{j-1}=0,
\qquad j=1,2,\dots\nn
\eea
where $\alpha_j$, $\beta_j$ and $\gamma_j$ are simple functions of the
frequency $\omega$, $l$ and $s$. Their explicit form can be found in
\cite{L}. The quasinormal mode boundary conditions are satisfied when
the following continued-fraction condition on the recursion
coefficients holds:
\be\label{CF}
0=\beta_0-
{\alpha_0\gamma_1\over \beta_1-}
{\alpha_1\gamma_2\over \beta_2-}\dots
\ee
The $n$--th quasinormal frequency is (numerically) the most stable
root of the $n$--th inversion of the continued-fraction relation
(\ref{CF}), i.e., it is the root of
\be\label{CFI}
\beta_n-
{\alpha_{n-1}\gamma_{n}\over \beta_{n-1}-}
{\alpha_{n-2}\gamma_{n-1}\over \beta_{n-2}-}\dots
{\alpha_{0}\gamma_{1}\over \beta_{0}}=
{\alpha_n\gamma_{n+1}\over \beta_{n+1}-}
{\alpha_{n+1}\gamma_{n+2}\over \beta_{n+2}-}\dots
\qquad (n=1,2,\dots).\nn
\ee
The infinite continued fraction appearing in equation (\ref{CFI}) can
be summed ``bottom to top'' starting from some large truncation index
$N$. Nollert \cite{N} has shown that the convergence of the procedure
improves if the sum is started using a wise choice for the value of
the ``rest'' of the continued fraction, $R_N$, defined by the relation
\be
R_N={\gamma_{N+1}\over\beta_{N+1}-\alpha_{N+1}R_{N+1}}\,.
\ee
Assuming that the rest can be expanded in a series of the form
\be\label{RN}
R_N=\sum_{k=0}^{\infty}C_k N^{-k/2}\,,
\ee
it turns out that the first few coefficients in the series are
$C_0=-1$, $C_1=\pm\sqrt{-2\ii \omega}$, $C_2=(3/4+2\ii \omega)$ and
$C_3=\left[l(l+1)/2+2\omega^2+3\ii\omega/2+3/32\right]/C_1$ (the
latter coefficient contains a typo in \cite{N}, but it is numerically
irrelevant anyway).

\subsection{Results}

\begin{table}
\centering
\caption{
Representative Schwarzschild quasinormal frequencies for $l=2$ and
$l=3$ (from \cite{L}).
}
\vskip 12pt
\begin{tabular}{@{}lll@{}}
\hline
\hline
  &$l=2$                 & $l=3$     \\
\hline
n &$\omega_n$            & $\omega_n$\\
\hline
1 &(0.747343,-0.177925)  &(1.198887,-0.185406)\\
2 &(0.693422,-0.547830)  &(1.165288,-0.562596)\\
3 &(0.602107,-0.956554)	 &(1.103370,-0.958186)\\
4 &(0.503010,-1.410296)	 &(1.023924,-1.380674)\\
5 &(0.415029,-1.893690)	 &(0.940348,-1.831299)\\
6 &(0.338599,-2.391216)	 &(0.862773,-2.304303)\\
7 &(0.266505,-2.895822)	 &(0.795319,-2.791824)\\
8 &(0.185617,-3.407676)	 &(0.737985,-3.287689)\\
9 &(0.000000,-3.998000)	 &(0.689237,-3.788066)\\
10&(0.126527,-4.605289)	 &(0.647366,-4.290798)\\
11&(0.153107,-5.121653)	 &(0.610922,-4.794709)\\
12&(0.165196,-5.630885)	 &(0.578768,-5.299159)\\
20&(0.175608,-9.660879)	 &(0.404157,-9.333121)\\
30&(0.165814,-14.677118) &(0.257431,-14.363580)\\
40&(0.156368,-19.684873) &(0.075298,-19.415545)\\
41&(0.154912,-20.188298) &(-0.000259,-20.015653)\\
42&(0.156392,-20.685630) &(0.017662,-20.566075)\\
50&(0.151216,-24.693716) &(0.134153,-24.119329)\\
60&(0.148484,-29.696417) &(0.163614,-29.135345)\\
\hline
\hline
\end{tabular}
\label{tab1}
\end{table}

\begin{figure*}
\begin{center}
\epsfig{file=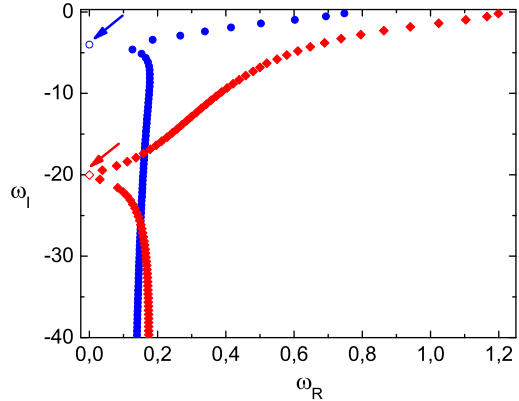,width=8cm,angle=0}
\caption{Quasinormal frequencies for gravitational perturbations with
$l=2$ (blue circles) and $l=3$ (red diamonds). Compare eg. Figure 1 in
\cite{N}. In both cases we mark by an arrow the algebraically special
mode, that is given analytically by Equation (\ref{AlgSp}); a more
extensive discussion of this mode is given in section
\ref{sec:kerras}. Notice that as the imaginary part of the frequency
tends to infinity the real part tends to a finite, {\it
$l$-independent} limit.
\label{fig:fig1}}
\end{center}
\end{figure*}

We computed quasinormal frequencies using Leaver's technique as
improved by Nollert. The slowly damped modes of the resulting spectrum
for $l=2$ and $l=3$ are shown in Table \ref{tab1} and Figure
\ref{fig:fig1}. Reintroducing physical units, the fundamental
oscillation frequency $f=\omega_R/(2\pi)$ and the damping time
$\tau=1/|\omega_I|$ of an astrophysical black hole scale with mass
according to the relation
\be
f=1.207 \left(\frac{10~M_\odot}{M}\right) {\rm kHz}\,,\qquad
\tau=0.5537 \left(\frac{M}{10~M_\odot}\right) {\rm ms}\,.
\ee
An ``algebraically special'' mode, whose frequency is (almost) pure
imaginary, separates the lower quasinormal mode branch from the upper
branch \cite{Cas}. This algebraically special mode has a very peculiar
nature, as we will see in section \ref{sec:kerras}. It is located at
\be\label{AlgSp}
\tilde \Omega_l=\pm \ii{(l-1)l(l+1)(l+2)\over 6}\,,
\ee
and it can be taken as roughly marking the onset of the asymptotic
high-damping regime. The algebraically special mode quickly moves
downwards in the complex plane as $l$ increases: from Table \ref{tab1}
we see that it corresponds to an overtone index $n=9$ when $l=2$, and
to an overtone index $n=41$ when $l=3$. This means that for high
values of $l$ the asymptotic high-damping regime sets in later,
becoming harder to probe using numerical methods.

Nollert was able to compute for the first time highly damped
quasinormal frequencies corresponding to {\it gravitational}
perturbations \cite{N}. His main result was that the real parts of the
quasinormal frequencies are well fitted, for large $n$, by a relation
of the form
\be\label{fit}
\omega_R=\omega_{\infty}+{\lambda_{s,l}\over \sqrt{n}}\,.
\ee
The leading-order fitting coefficient $\omega_{\infty}=0.0874247$ is
independent of $l$. Corrections of order $\sim n^{-1/2}$, however, are
$l$-dependent (we will see in a moment that they also depend on the
spin $s$ of the perturbing field; that's why we denoted them by
$\lambda_{s,l}$).  For gravitational perturbations ($s=-2$) Nollert
found that $\lambda_{-2,2}=0.4850$, $\lambda_{-2,3}=1.067$,
$\lambda_{-2,6}=3.97$. Numerical data for scalar perturbations ($s=0$)
are also well fitted by formula (\ref{fit}), and the asymptotic
frequency for scalar modes is exactly the same. Only the leading-order
correction coefficients $\lambda_{0,l}$ are different:
$\lambda_{0,0}=0.0970$, $\lambda_{0,1}=0.679$, $\lambda_{0,2}=1.85$
\cite{BK}.

These numerical results are perfectly consistent with analytical
calculations. Motl \cite{M} analyzed the continued fraction condition
(\ref{CF}) to find that highly damped quasinormal frequencies satisfy
the relation
\be\label{Mresult}
\omega\sim{T_H \ln 3}+(2n+1)\pi \ii T_H+{\cal O}(n^{-1/2})\,.
\ee
(remember that in our units $2M=1$, so that the Hawking temperature of
a Schwarzschild black hole $T_H=1/4\pi$). This conclusion was later
confirmed by complex-integration techniques \cite{MN} and
phase-integral methods \cite{AH}. Neitzke \cite{N2} first suggested
that leading order corrections to the asymptotic frequency should be
proportional to $[(s^2-1)-3l(l+1)]$; this is consistent with the
numerical values of the $\lambda_{s,l}$'s. The proportionality
constant is dependent on $s$: $\lambda_{s,l}=k_s[(s^2-1)-3l(l+1)]$.
Maassen van den Brink \cite{MVDBas} derived $k_{-2}$ analytically,
finding
\be\label{km2}
k_{-2}=-{\sqrt{2}[\Gamma(1/4)]^4\over 432 \pi^{5/2}}\simeq -0.0323356\,,
\ee
in perfect agreement with the numerical data, that yield
$k_{-2}=-0.0323$, $k_{0}=-0.0970=3k_{-2}$ \cite{BK}.

Finally, a systematic perturbative approach to determine lower-order
corrections to formula (\ref{fit}) has been developed in \cite{ms}.
Their final result, valid for scalar and gravitational perturbations,
is again perfectly consistent with numerical data. The perturbative
technique used in \cite{ms} has also been extended to
higher-dimensional Schwarzschild-Tangherlini black holes with the
result \cite{CLY}
\be\label{CLYf}
\omega=T_H \ln (1+2\cos\pi j)+(2n+1)\pi\ii T_H
+k_D \omega_I^{-(D-3)/(D-2)}\,,
\ee
where the coefficient $k_D$ can be determined analytically for given
values of $D$ and $j$ \cite{highd}.

Leaver's method can easily be generalized to these higher-dimensional
black holes \cite{CLY}; the recursion relation has four terms for
vector-gravitational and tensor-gravitational perturbations (see also
\cite{highdWKB} for similar calculations using WKB methods). The
most impressive example of an application of Leaver's technique to
date is probably reference \cite{CLY2}. There the continued fraction
method has been used for a detailed calculation of
scalar-gravitational quasinormal frequencies in the case of
5-dimensional Schwarzschild black holes using an {\it eight-term}
recursion relation.

In conclusion, all numerical and analytical results for non-rotating
black holes are in perfect agreement. As $|\omega_I|\to \infty$ scalar
and gravitational oscillation frequencies of non-rotating
Schwarzschild-Tangherlini black holes {\it in all dimensions} tend to
the constant value $\omega_R=T_H \ln 3$ (this was first argued in
\cite{MN} and explicitly shown in \cite{birm}). For electromagnetic
perturbations the situation is less clear: analytical and numerical
results in $D=4$ suggest that the asymptotic limit should be
$\omega_R=0$. For electromagnetic perturbations in $D=5$, formula
(\ref{CLYf}) doesn't even make sense: this probably means that there
are no asymptotic quasinormal frequencies at all (a possibility first
suggested in \cite{MN}). A similar behavior occurs for the
3--dimensional ``draining bathtub'' metric describing a rotating
acoustic black hole \cite{BCL}, suggesting that this feature may
somehow be related to odd-dimensional spacetimes.

%%%%%%%%%%%%%%%%%%%%%%%%%%%%%%%%%%%%%%%%%%%%%%%%%%%%%%%%%%%%%%%%%%%%%%%%%%%%%%%
%%%%%%%%%%%%%%%%%%%%%%%%%%%%%%%%%%%%%%%%%%%%%%%%%%%%%%%%%%%%%%%%%%%%%%%%%%%%%%%

\section{Reissner-Nordstr\"om black holes}
\label{sec:rn}

\subsection{Computational method}

The computational procedure described in the previous section can be
generalized to RN black holes. Here we only sketch the key steps,
referring the reader to \cite{L2} for more details. As usual we define
a tortoise coordinate $r_*$ by the relation (\ref{tortoise}), but now
$\Delta=r^2-r+Q^2$ (recall that in our units $0\leq Q\leq
1/2$). Explicitly, the tortoise coordinate can be written as
\be
r_*=r+\frac{r_+^2}{r_+-r_-}\ln(r-r_+)-\frac{r_-^2}{r_+-r_-}\ln(r-r_-)\,,
\ee 
where $r_\pm=(1\pm\sqrt{1-4Q^2})/2$ is the location of the inner
(Cauchy) and outer (event) horizons of the RN metric. After separation
of the angular dependence and Fourier decomposition, electromagnetic
and axial--gravitational perturbations of the RN metric are described
by two wave equations,
\be\label{axialRN}
\left({d^2\over dr_*^2}+\omega^2\right)Z_i^-=V_i^-Z_i^-\,,
\ee
where
\be\label{RNpot}
V_i^{-}=\frac{\Delta}{r^5}\left[l(l+1)r-q_j+\frac{4Q^2}{r}\right]\,,\qquad
i,j=1,2~(i\neq j)\,,
\ee
and
\be
q_1=\left[3+\sqrt{9+16Q^2(l-1)(l+2)}\right]/2\,,\qquad
q_2=\left[3-\sqrt{9+16Q^2(l-1)(l+2)}\right]/2\,.
\ee
As in the Schwarzschild case, the polar perturbation variables
($Z_i^+$) can be obtained from the axial variables ($Z_i^-$) by a
Chandrasekhar transformation \cite{MTB}, so we do not consider them in
the following. In the limit $Q\to 0$ the potentials $V_1^-$ and
$V_2^-$ describe, respectively, purely electromagnetic [spin weight
$s=-1$ in Equation (\ref{sax})] and pure axial--gravitational [spin
weight $s=-2$ in Equation (\ref{sax})] perturbations of a
Schwarzschild black hole; but for any non-zero value of the charge,
electromagnetic and gravitational perturbations are tangled. The
perturbation equations are solved by a series expansion of the form
\be\label{RNser}
Z_i^{-}=\frac{r_+e^{-2\ii\omega r_+}(r_+-r_-)^{-2\ii\omega-1}
(r-r_-)^{1+\ii\omega}e^{\ii\omega r}}{r}u^{-\ii\omega r_+^2/(r_+-r_-)}
\sum_{j=0}^\infty a_j u^j\,,
\ee
where $u=(r-r_+)/(r-r_-)$. The coefficients $a_j$ of the expansion are
now determined by a four-term recursion relation whose coefficients
depend on the angular index $l$, on the frequency $\omega$ and on the
charge $Q$. The problem can be reduced to a three-term recursion
relation of the form (\ref{CF}) using a Gaussian elimination step (see
\cite{L2} for details). Then we can use again a continued-fraction
condition to determine quasinormal frequencies, and Nollert's
procedure can be applied to accelerate convergence for large values of
the damping. The first few coefficients in the series (\ref{RN}) are
now $C_0=-1$, $C_1=\pm\sqrt{-2\ii\omega (2r_+-1)}$,
$C_2=(3/4+2\ii\omega r_+)$.

Due to convergence reasons, continued fraction calculations become
more and more computationally intensive for large values of the charge
\cite{L2}. In the limit of maximally charged black holes ($Q=1/2$) the
wave equations have a different singularity structure, and deserve a
special treatment \cite{OMOI}. We discuss the peculiarities of the
extremal RN case in section \ref{sec:rnse}.

\subsection{Results}

The first few overtones of a RN black hole were studied numerically by
Andersson and Onozawa \cite{A2,AO} using different numerical methods
\cite{modeorder}.  These studies revealed a very peculiar behavior.

\begin{figure}[h]
\begin{center}
\epsfig{file=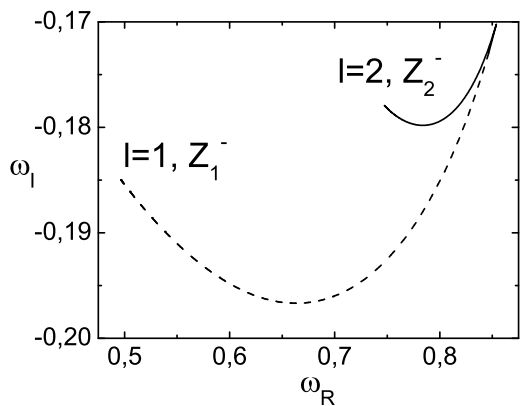,width=8cm,angle=0}
\caption{Trajectory described in the complex--frequency plane by the
fundamental RN quasinormal mode as the charge is increased. The solid
line corresponds to $l=2$ and $Z_2^-$ (perturbations that reduce to
the axial--gravitational Schwarzschild case as $Q\to 0$); the dashed
line, to $l=1$ and $Z_1^-$ (purely electromagnetic Schwarzschild
perturbations in the limit $Q\to 0$). The modes coalesce in the
extremal limit $Q\to 1/2$.}
\label{fig:figrn}
\end{center}
\end{figure}

The solid line in Figure \ref{fig:figrn} is the trajectory described
in the complex--frequency plane by the fundamental quasinormal mode
with $l=2$ corresponding to $Z_2^-$ (perturbations that reduce to the
axial--gravitational Schwarzschild case). The dashed line is the same
trajectory for the fundamental quasinormal mode with $l=1$
corresponding to $Z_1^-$ (which limits us to purely electromagnetic
perturbations). The Schwarzschild limit corresponds to the bottom left
of each curve, and the trajectories are described counterclockwise as
$Q$ increases. The real part of the frequency grows monotonically with
$Q$, and the imaginary part shows a relative minimum. This behavior
can easily be understood using WKB arguments: restoring the mass in
the equations, the lowest order WKB approximation yields
\be
\omega_R\sim \left(l+\frac{1}{2}\right)
\sqrt{\frac{(M-Q^2/r_0)}{r_0^3}}\,,\qquad
\omega_I\sim -\frac{1}{2}
\sqrt{\frac{(M-Q^2/r_0)(3M-4Q^2/r_0)}{r_0^4}}\,,
\ee
where $r_0\sim (3M+\sqrt{9M^2-8Q^2})/2$ is the location of
the potential peak. It is worth noting that {\it modes of $Z_2^-$ with
angular index $l$ coalesce with modes of $Z_1^-$ with index $(l-1)$ in
the extremal limit}. This is a general feature we are going to discuss
in section \ref{sec:rnse}. 

\begin{figure*}
\begin{center}
\begin{tabular}{cc}
\epsfig{file=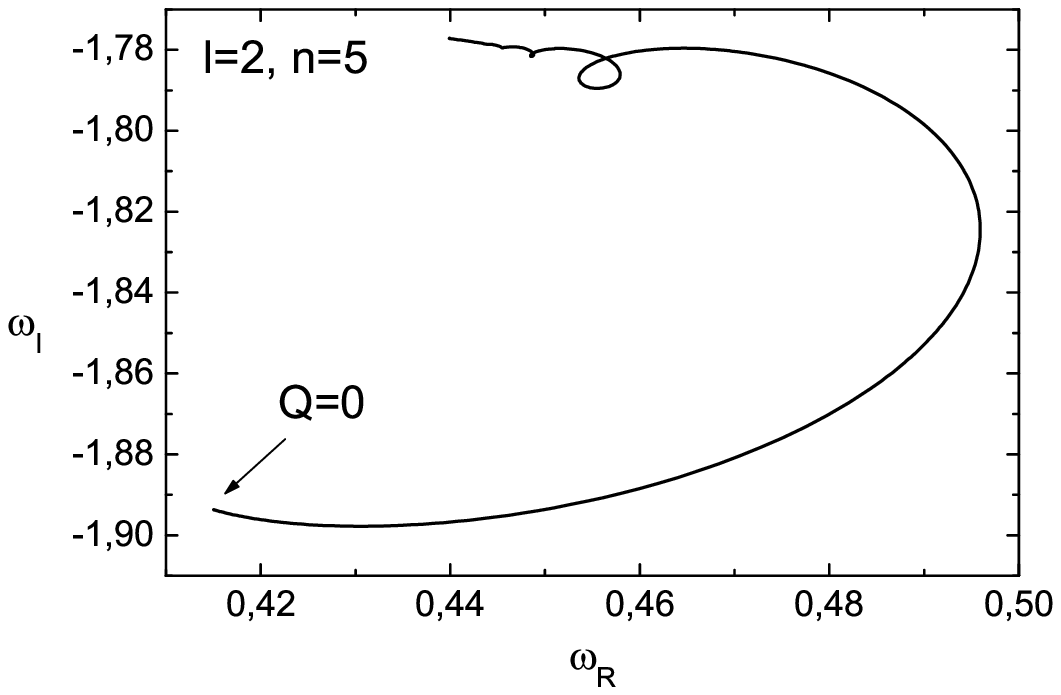,width=8cm,angle=0} &
\epsfig{file=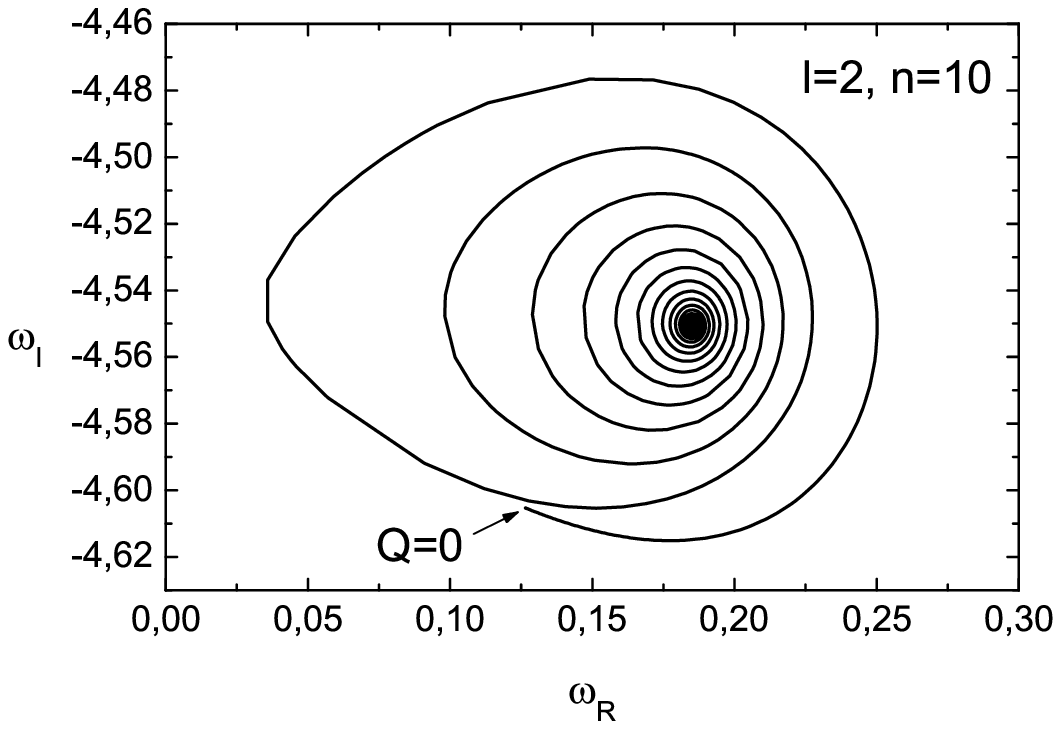,width=8cm,angle=0} \\
\epsfig{file=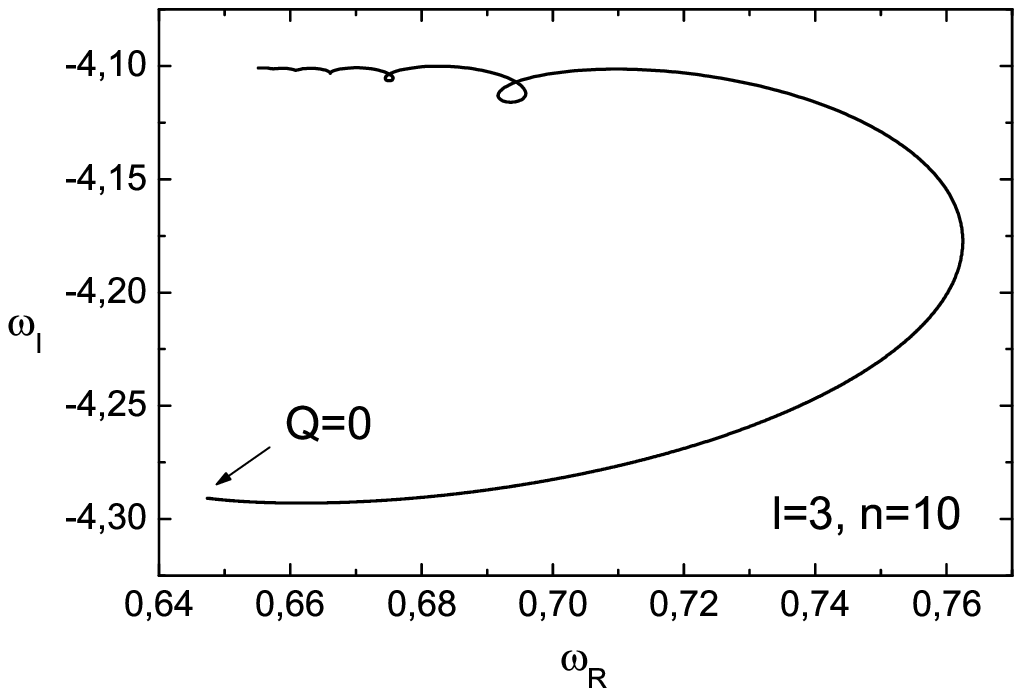,width=8cm,angle=0} &
\epsfig{file=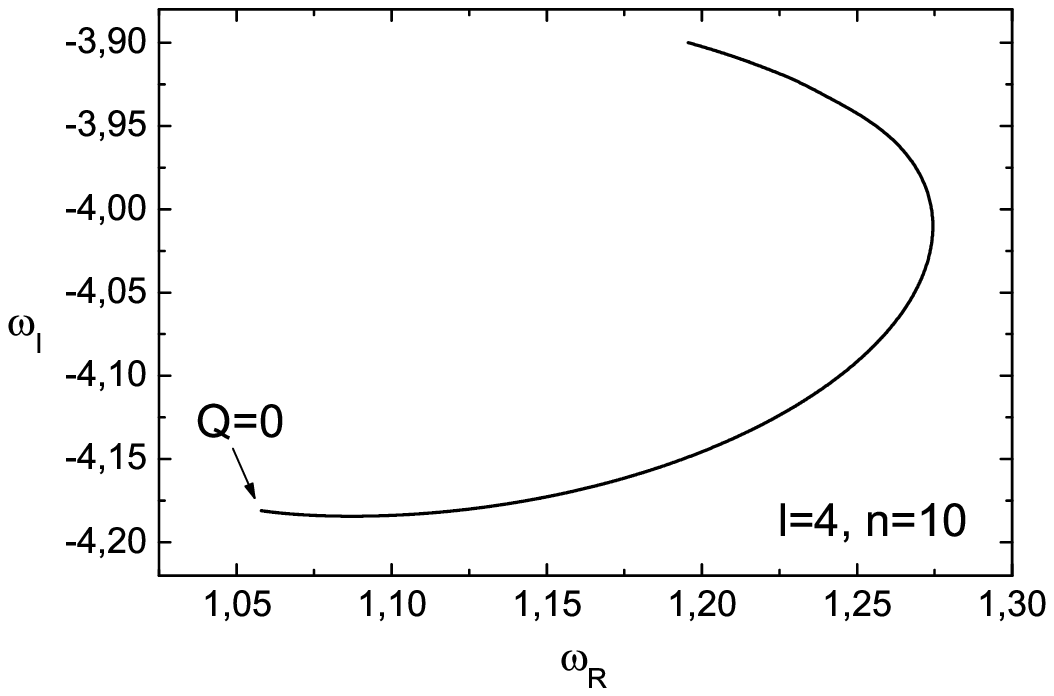,width=8cm,angle=0}
\end{tabular}
\caption{The top two panels show the behavior of the $n=5$ and $n=10$
quasinormal frequencies in the complex $\omega$ plane. The $n=10$ mode
``spirals in'' towards its value in the extremal charge limit; the
number of spirals described by each mode increases roughly as the mode
order $n$. The panels in the second row show how the $n=10$ spiral
``unwinds'' as the angular index $l$ is increased (in other words, the
asymptotic behavior sets in later for larger $l$'s). In all cases, we
have marked by an arrow the frequency corresponding to the
Schwarzschild limit ($Q=0$).
\label{fig:fig2}}
\end{center}
\end{figure*}

The algebraically special frequency of Reissner-Nordstr\"om black
holes can be found using the same reasoning as in the Schwarzschild
case \cite{Cas}. It reads
\be\label{RNas}
\tilde \Omega_l^{(i)}=\pm \ii
\frac{(l-1)l(l+1)(l+2)}{3\left[1\pm \sqrt{1+4 Q^2(l-1)(l+2)}\right]}\,,
\ee
where the superscript $(i)$ refers to the two different potentials in
Equation (\ref{axialRN}). For extremal black holes ($Q=1/2$) the first
algebraically special mode for $l=2$ is located at $\Omega=-3\ii$.

\begin{figure}[h]
\begin{center}
\begin{tabular}{cc}
\epsfig{file=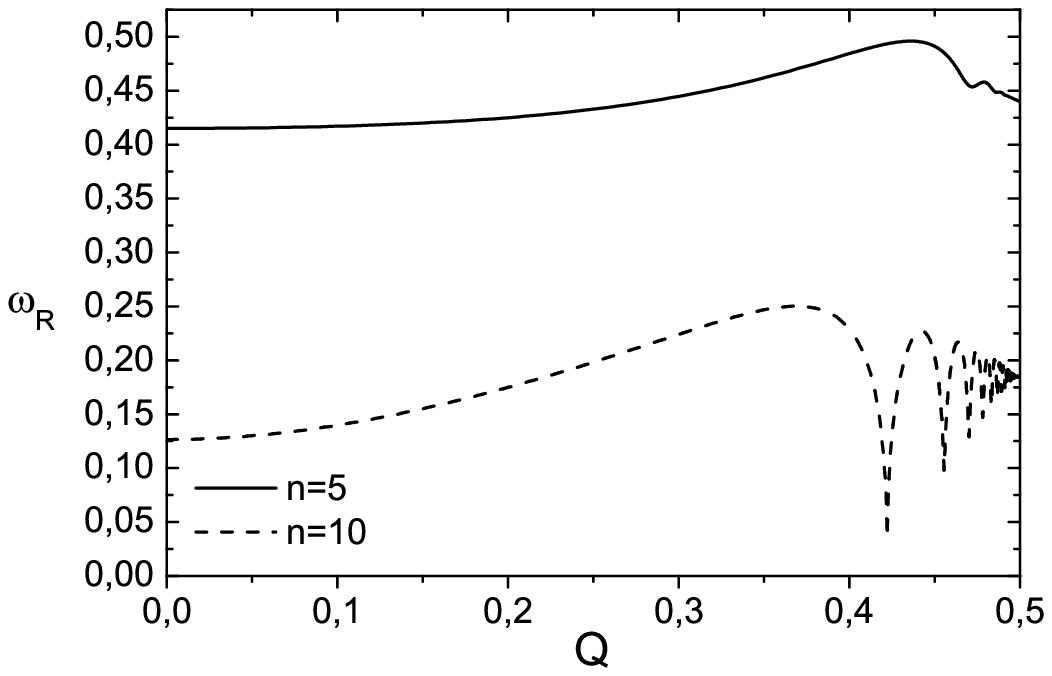,width=8cm,angle=0} &
\epsfig{file=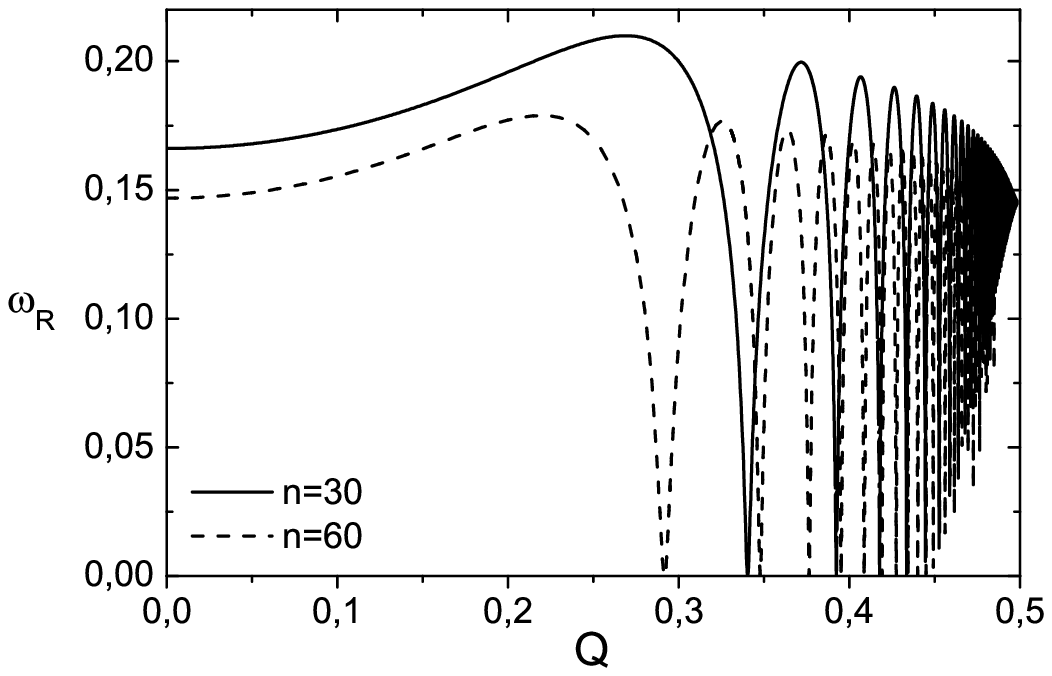,width=8cm,angle=0} \\
\epsfig{file=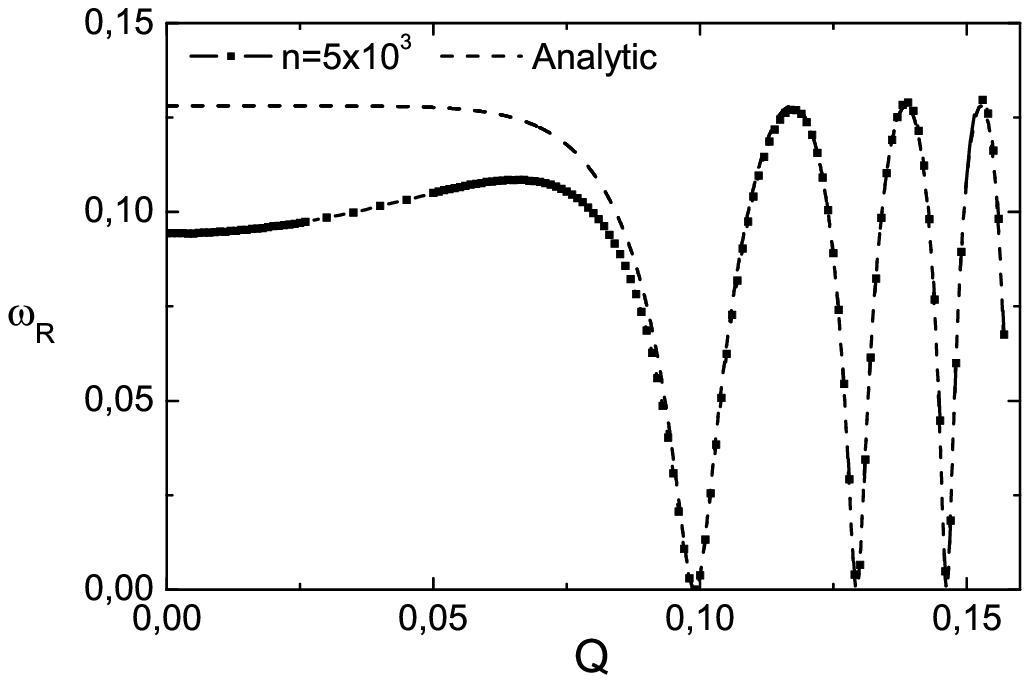,width=8cm,angle=0} &
\epsfig{file=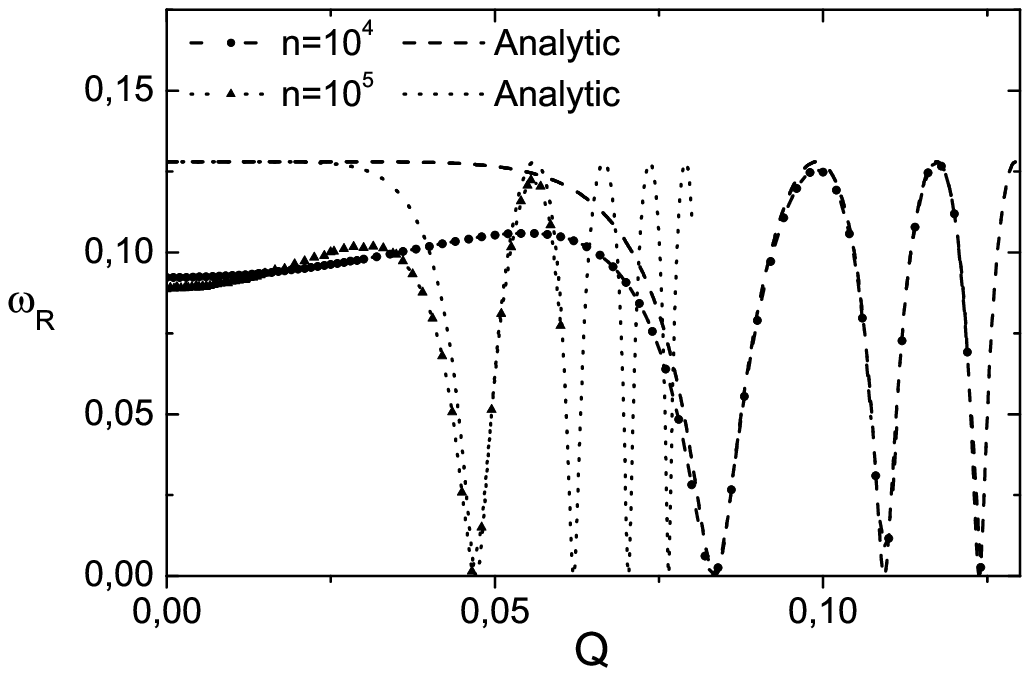,width=8cm,angle=0}
\end{tabular}
\caption{Real part of the RN quasinormal frequencies as a function of
charge for $n=5,~10,~30,~60,~5000,~10000,~100000$. As the mode order
increases the computation becomes more and more time consuming, the
oscillations become faster, and a good numerical sampling is rather
difficult to achieve; therefore in the last plot we use different
symbols (small squares, circles and triangles) to display the actually
computed points. For $n=5000,~10000,~100000$ we also compare to the
prediction of the analytic formula (\ref{MNf}) derived by Motl and
Neitzke \cite{MN}. The oscillatory behavior is reproduced extremely
well by their formula, but the disagreement increases for small
charge: formula (\ref{MNf}) does not yield $T_H \ln 3$ in the
Schwarzschild limit.}
\label{fig:fig3}
\end{center}
\end{figure}

\begin{figure*}
\begin{center}
\begin{tabular}{cc}
\epsfig{file=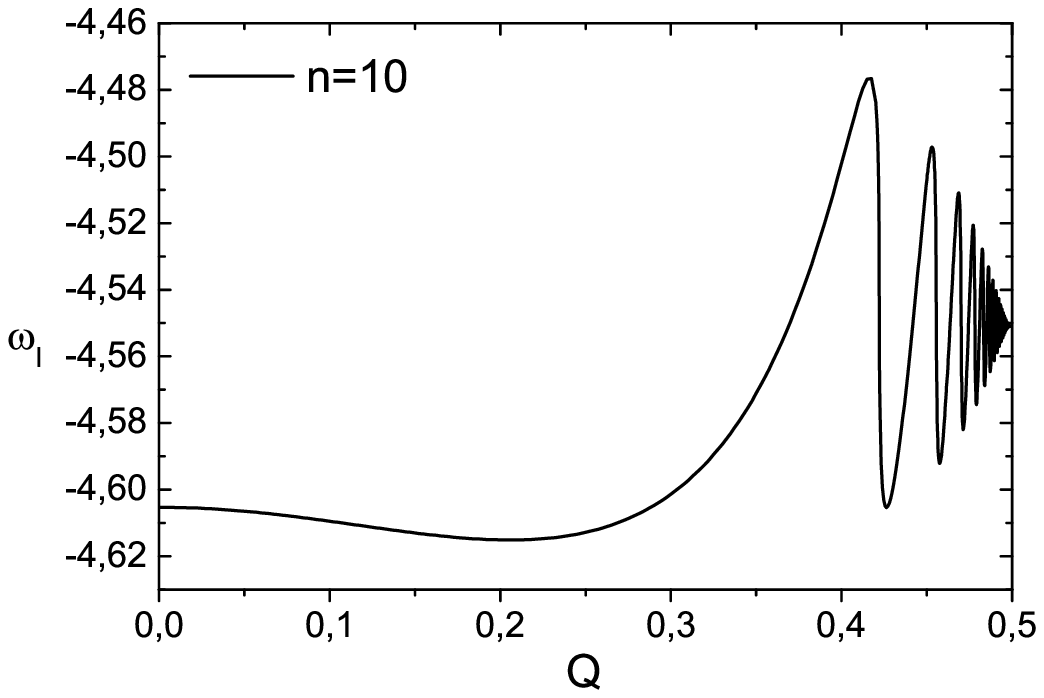,width=8cm,angle=0} &
\epsfig{file=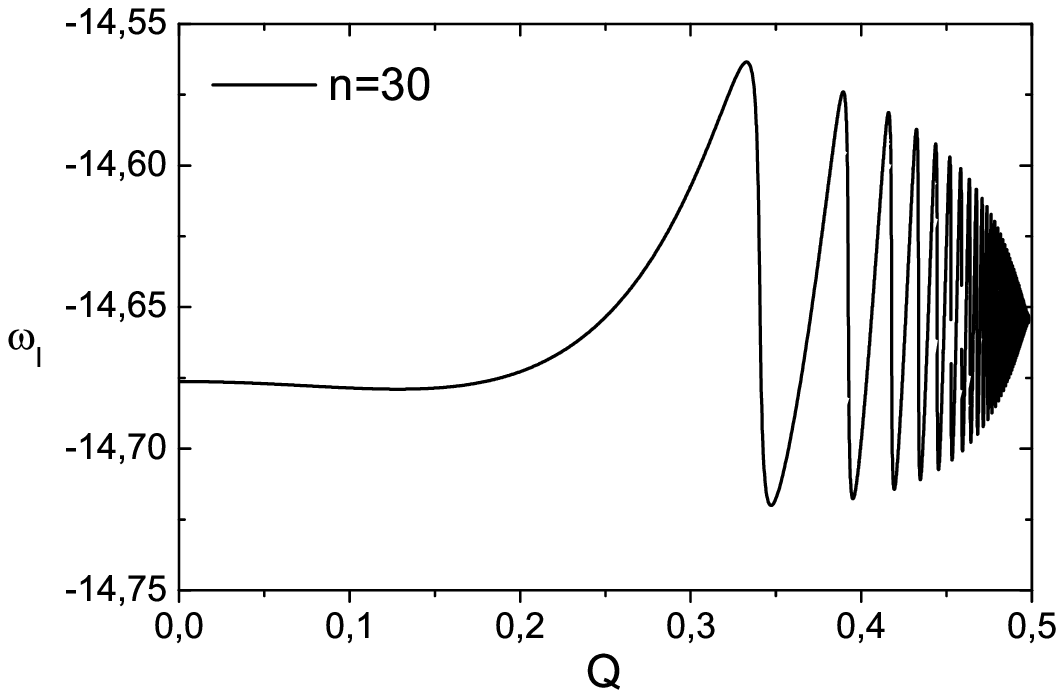,width=8cm,angle=0} \\
\epsfig{file=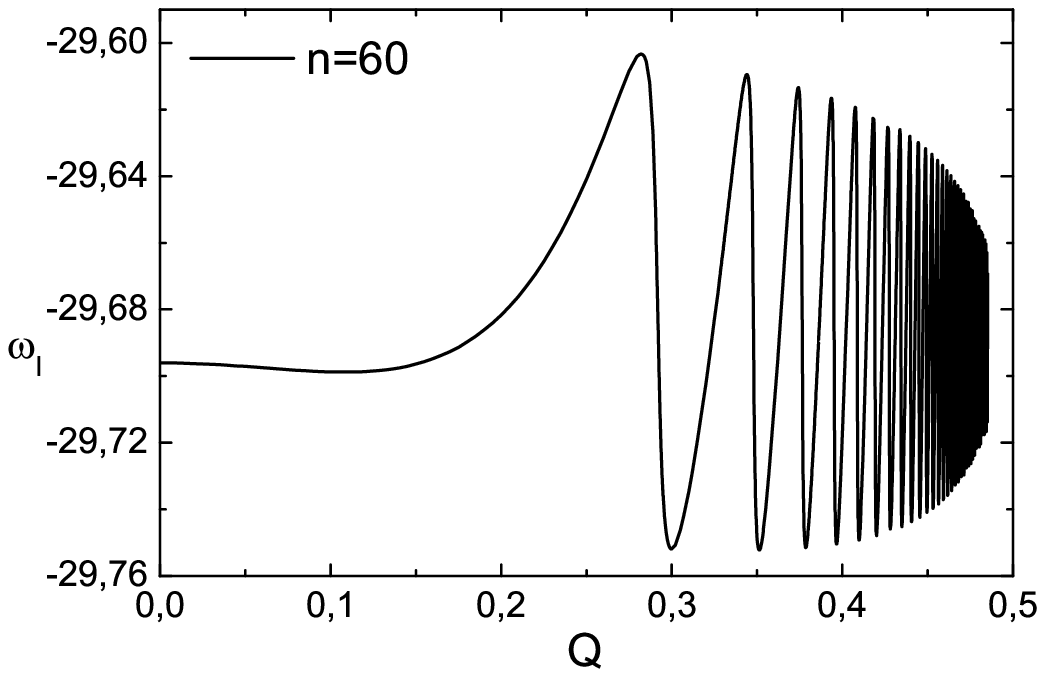,width=8cm,angle=0} &
\epsfig{file=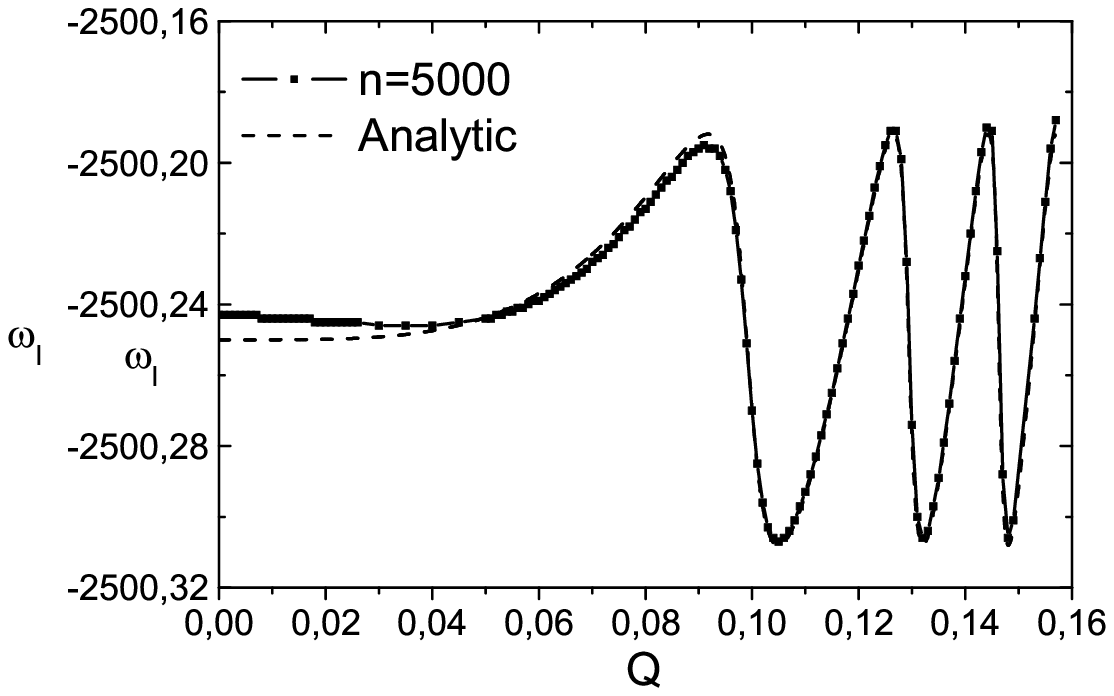,width=8cm,angle=0}
\end{tabular}
\caption{Imaginary part of the RN quasinormal frequencies as a
function of charge for $n=10,~30,~60,~5000$. For $n=5000$ we also
display the actually computed points, and compare to the prediction of
the analytic formula (\ref{MNf}). As for the real part, the
oscillations are reproduced extremely well, but the disagreement with
our numerical data increases for small charge.
\label{fig:fig4}}
\end{center}
\end{figure*}

In the rest of this section we explore the high-damping regime. We fix
our attention on the wave equation for $Z_2^-$ (perturbations reducing
to gravitational perturbations of Schwarzschild as $Q\to 0$); results
for $Z_1^-$ are similar. As can be seen in Figures \ref{fig:fig2},
\ref{fig:fig3} and \ref{fig:fig4}, the behavior of high order modes is
quite surprising. At first the trajectories described by the modes in
the complex-$\omega$ plane show closed loops, as in the top left panel
of Figure \ref{fig:fig2}. Then they get a spiral-like shape, moving
out of their Schwarzschild value and looping in towards some limiting
frequency as $Q$ tends to the extremal value. This kind of behavior is
shown in the top right panel of Figure \ref{fig:fig2}. We observe that
such a spiralling behavior sets in for larger values of the modes'
imaginary part (i.e., larger values of $n$) as the angular index $l$
increases. In other words, increasing $l$ for a given value of the
mode index $n$ has the effect of unwinding the spirals. This can be
seen in the two bottom panels of Figure \ref{fig:fig2}. However, for
each $l$ the spiralling behavior is eventually observed when $n$ is
large enough.

A perhaps clearer picture of the modes' behavior can be obtained
looking separately at the real and imaginary parts of the mode
frequencies as function of charge. Let us focus first on the real
parts. The corresponding numerical results are shown in Figure
\ref{fig:fig3}.  It is quite apparent that, as the mode order grows,
the oscillating behavior as a function of charge start earlier and
earlier. For large $n$ the oscillations become faster, the convergence
of the continued fraction method slower, and the required computing
time gets longer. Therefore, when the imaginary part increases it
becomes more difficult to follow the roots numerically as we approach
the extremal value $Q=1/2$. That's why data for very large values of
$n$ do not cover the whole range of allowed values for $Q$. Despite
these difficulties, we have many reasons to trust our numerics. We
have carefully checked our results, using first double and then
quadrupole precision in our Fortran codes (indeed, as $n$ increases,
we can obtain results for large values of the charge only using
quadrupole precision). Our frequencies accurately reproduce Nollert's
results in the Schwarzschild limit, so the numerics can be trusted for
small values of $Q$. More importantly, at large values of the charge
numerical data are in excellent agreement with analytical predictions
\cite{MN}. Motl and Neitzke found the following result for the
asymptotic frequencies of scalar and electromagnetic-gravitational
perturbations of a RN black hole \cite{unitsMN}:
\be\label{MNf}
e^{\beta \omega}+2+3e^{-\beta_I \omega}=0.
\ee
This implicit formula for the asymptotic quasinormal modes was later
rederived by an indepent method in \cite{AH}. The complex solutions of
Equation (\ref{MNf}) exactly overlap with the oscillations we observe
at large damping for large enough values of $Q$ (see Figures
\ref{fig:fig3} and \ref{fig:fig4}). This agreement gives support to
the asymptotic formula (\ref{MNf}) and confirms that the numerical
calculation is still accurate for large values of $Q$ \cite{N2}.

Similar considerations apply to the imaginary parts. Some plots
illustrating the general trend are shown in Figure
\ref{fig:fig4}. Once again the numerics show excellent agreement with
the asymptotic formula (\ref{MNf}) as $n$ increases. The analytic
formula deviates from numerical results only for small values of the
charge.

In conclusion, numerical results lend strong support to the analytical
expression (\ref{MNf}). This formula presents us with some striking
puzzles. As emphasized in \cite{MN,N2}, the predicted asymptotic RN
quasinormal frequencies do not reduce to the expected Schwarzschild
limit, Equation (\ref{Mresult}). One finds instead
\be
\omega_R\to T_H \ln 5 \qquad \text{as} \qquad Q\to 0\,,
\ee
a prediction that is hard to reconcile with Hod's interpretation of
the asymptotic quasinormal spectrum.  Surprisingly, in the extremal
limit $Q\to 1/2$ the real part of the frequency predicted by Equation
(\ref{MNf}) coincides with the Schwarzschild value:
\be\label{xrnl}
\omega_R\to T_H \ln 3 \qquad \text{as} \qquad Q\to 1/2\,.
\ee
It is not clear to which extent this result is relevant: in the
extremal limit the two horizons coalesce, the topology of the
singularities in the differential equation changes, and the problem
may require a separate analysis. Indeed, there are some arguments that
the asymptotic oscillation frequency for extremal RN black holes is
{\it not} given by Equation (\ref{xrnl}) \cite{das}.

Another bizarre prediction is that -- according to Equation
(\ref{MNf}) -- asymptotic quasinormal frequencies of charged black
holes depend not only on the black hole's Hawking temperature, but
also on the Hawking temperature of the (causally disconnected) inner
horizon. The dependence on the causally disconnected region is not
surprising when one looks at the calculation by Motl and Neitzke:
their result depends only on the behavior of the potential close to
the curvature singularity. Still, the region inside the event horizon
plays no role whatsoever in the classical analysis of black hole
perturbations. From this perspective, the appearance of $\beta_I$ in
Equation (\ref{MNf}) is somehow disturbing.

An interesting classification of the solutions of Equation (\ref{MNf})
can be found in Figure 3 and section 4.4 of \cite{AH}. Besides
``spiralling'' solutions the equation also admits {\it periodic}
solutions when $\kappa\equiv \sqrt{1-Q^2/M^2}$ is a rational number,
and even {\it pure imaginary} solutions that may not be quasinormal
modes at all!

To conclude this section: analytical and numerical calculations
indicate that (\ref{MNf}) is correct. Still, we are far from a full
understanding of the physical meaning of this equation and its
relation with the asymptotic Schwarzschild quasinormal spectrum
(\ref{Mresult}).

%%%%%%%%%%%%%%%%%%%%%%%%%%%%%%%%%%%%%%%%%%%%%%%%%%%%%%%%%%%%%%%%%%%%%%%%%%%%%%%
%%%%%%%%%%%%%%%%%%%%%%%%%%%%%%%%%%%%%%%%%%%%%%%%%%%%%%%%%%%%%%%%%%%%%%%%%%%%%%%

\subsection{Extremal Reissner-Nordstr\"om black holes}
\label{sec:rnse}

Leaver's method fails in the extremal limit. In this limit the two
horizons coalesce ($r_{\rm hor}=r_+=r_-=1/2$). The tortoise coordinate
becomes, in our units,
\be
r_*=r+\ln(2r-1)-\frac{1}{4r-2}\,,
\ee 
and the radial wave equation has irregular singular points at the
horizon and at infinity. The series expansion (\ref{RNser}) is not
valid any more. The trick used by Onozawa {\it et al.} \cite{OMOI} is
to change the expansion variable: they expand in $u=(r-2r_{\rm
hor})/r$. This choice pushes the horizon to $u=-1$ and infinity to
$u=1$.  The requirement that a series solution of the form $\sum_j a_j
u^j$ converges at both boundaries means that $\sum_j a_j$ and $\sum_j
(-1)^j a_j$ should both be finite; that is, that both $\sum_j a_{2j}$
and $\sum_j a_{2j+1}$ must be convergent. Substitution of the series
$\sum_j a_j u^j$ in the differential equation yields a five-term
recurrence relation. This five-term recurrence relation can be split
in {\it two} separate five-term recurrence relations, corresponding to
the convergence conditions for even and odd coefficients. Finally, the
even and odd five-term recurrence relations are reduced to three terms
by two iterations of a Gaussian elimination step.

\begin{figure*}
\begin{center}
\epsfig{file=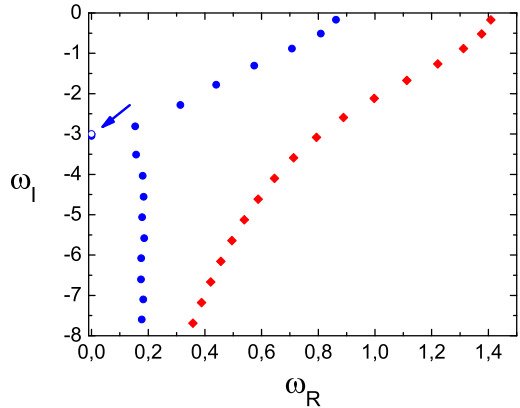,width=8cm,angle=0}
\caption{Quasinormal frequencies for gravitational--type perturbations
of an extremal RN black hole with $l=2$ (blue circles) and $l=3$ (red
diamonds). The spectrum for electromagnetic--type perturbations with
$l=1$ and $l=2$ would be exactly the same. We mark by an arrow the
location of the algebraically special mode. We find a quasinormal mode
at $\omega=(0,-3.047876)$ [filled blue circle], while Chandrasekhar's
formula (\ref{RNas}) predicts a mode at $\omega=(0,-3)$ [empty blue
circle]. Unfortunately the numerical method becomes unstable for
values of $|\omega_I|\gtrsim 10$, and cannot be used to verify the
prediction (\ref{xrnl}) of the Motl-Neitzke formula (\ref{MNf}).
\label{fig:figxrn}}
\end{center}
\end{figure*}
%The mode for $l=3$ should be located at $\omega=(0,-12)$.

A careful analysis of the extremal RN case is interesting for many
reasons. First of all, as we have anticipated (see Figure
\ref{fig:figrn}), the quasinormal mode spectrum for extremal RN black
holes is characterized by an isospectrality between electromagnetic
and gravitational perturbations: {\it modes of $Z_2^-$ with angular
index $l$ coalesce with modes of $Z_1^-$ with index $(l-1)$ in the
extremal limit}. This has been shown to be a manifestation of
supersymmetry \cite{OOMIKRW}. The supersymmetric nature of extremal
Reissner-Nordstr\"om black holes is an attractive feature for the
calculation of quantum effects, and it has been exploited in various
contexts. Furthermore, topological arguments have been used to show
that the entropy-area relation breaks down for extremal RN black holes
\cite{HHR}. So we believe that some caution is required in claiming
that the connection between quasinormal modes and the area spectrum is
still valid for extreme black holes, as recently advocated in
\cite{ACL}. These problems may be connected with the finding that
extremally charged black holes in a (non-asymptotically flat) anti-de
Sitter spacetime could in principle be marginally unstable
\cite{BKAdS}.

The explicit analysis of the extremal RN case confirms the above
picture \cite{OMOI}. It also shows that the standard Leaver method is
surprisingly accurate even in this ``extreme'' situation, where it is
not supposed to converge at all! Indeed, the largest error on the real
frequency of the fundamental gravitational mode introduced by the
``naive'' Leaver treatment is only $\sim 3~\%$. The authors of
\cite{OMOI} derived only a handful of quasinormal frequencies (namely
the first three overtones for different values of $l$). As far as we
know, results for extremal RN black holes in the intermediate damping
regime have never been published to date. We show the resulting
quasinormal spectrum in Figure \ref{fig:figxrn} (to be compared with
Figure \ref{fig:fig1}).  

Extremal RN quasinormal frequencies have a similar pattern to
Schwarzschild frequencies. Their real part first decreases,
approaching the pure-imaginary axis as the overtone index grows. We
find a quasinormal mode at $\omega=(0,-3.047876)$, while
Chandrasekhar's formula (\ref{RNas}) predicts a mode at
$\omega=(0,-3)$. Then $\omega_R$ increases again, apparently
approaching a constant value that could be compatible with the
prediction (\ref{xrnl}) of the Motl-Neitzke asymptotic formula
(\ref{MNf}). Unfortunately it is hard to get stable numerical results
for values of $|\omega_I|\gtrsim 10$ to explicitly verify this
prediction. This relative numerical instability is not unexpected,
since a) we need to solve simultaneously two different continued
fraction relations, and b) for each of them we must carry out two
Gaussian elimination steps.

%%%%%%%%%%%%%%%%%%%%%%%%%%%%%%%%%%%%%%%%%%%%%%%%%%%%%%%%%%%%%%%%%%%%%%%%%%%%%%%
%%%%%%%%%%%%%%%%%%%%%%%%%%%%%%%%%%%%%%%%%%%%%%%%%%%%%%%%%%%%%%%%%%%%%%%%%%%%%%%

\section{Kerr black holes}
\label{sec:kerr}

\subsection{Computational method}
\label{kcm}

As in the RN case, here we only give the basic steps of our
computational procedure. More details can be found in
\cite{N,L,O}. For Kerr black holes, the perturbation problem can be
reduced to two ordinary differential equations for the angular and
radial parts of the perturbations, respectively.  In Boyer-Lindquist
coordinates, defining $u=\cos\theta$, the angular equation reads
\be
\left[
(1-u^2)S_{lm,u}
\right]_{,u}
+\left[
(a\omega u)^2-2a\omega su+s+ _sA_{lm}-{(m+su)^2\over 1-u^2}
\right]
S_{lm}=0\,,
\label{angularwaveeq}
\ee
and the radial one
\be
\Delta R_{lm,rr}+(s+1)(2r-1)R_{lm,r}+V(r)R_{lm}=0\,,
\ee
where
\bea
V(r)&=&\left\{
\left[
(r^2+a^2)^2\omega^2-2am\omega r+a^2m^2+is(am(2r-1)-\omega(r^2-a^2))
\right]\Delta^{-1}\right.\nn
\\
&+&
\left.
\left[
2is\omega r-a^2\omega^2- _sA_{lm}
\right]
\right\}\,.
\eea
The parameter $s=0,-1,-2$ for scalar, electromagnetic and
gravitational perturbations respectively, $a$ is the Kerr rotation
parameter ($0\leq a\leq 1/2$), and $_sA_{lm}= _sA_{lm}(a\omega)$ is an
angular separation constant. In the Schwarzschild limit $a=0$ the
angular separation constant can be determined analytically, and is
given by the relation $_sA_{lm}=l(l+1)-s(s+1)$.

Boundary conditions for the angular and radial equations yield {\it
two} three-term continued fraction relations of the form (\ref{CF}).
Finding quasinormal frequencies is a two-step procedure: for assigned
values of $a,~\ell,~m$ and $\omega$, first find the angular separation
constant $_sA_{lm}(a\omega)$ looking for zeros of the {\it angular}
continued fraction; then replace the corresponding eigenvalue into the
{\it radial} continued fraction, and look for its zeros as a function
of $\omega$.  In principle, the convergence of the procedure for modes
with large imaginary parts can be improved, as described earlier, by a
wise choice of the rest, $R_N$, of the radial continued
fraction. Expanding this rest as in formula (\ref{RN}) and introducing
$b\equiv \sqrt{1-4a^2}$, we get for the first few coefficients:
$C_0=-1$, $C_1=\pm\sqrt{-2\ii\omega b}$, $C_2=\left[ 3/4+\ii\omega
(b+1)-s \right]$.

The angular continued fraction, considered {\it separately} from the
radial continued fraction, is a simple and efficient tool to compute
the angular eigenvalues $_sA_{lm}(a\omega)$ of the spin-weighted
spheroidal wave equations for complex values of the argument
$a\omega$. We simply start from the known Schwarzschild limit,
$_sA_{lm}(0)=l(l+1)-s(s+1)$. Then we solve the angular continued
fraction, using this value as an initial guess, to compute $_sA_{lm}$
for generic values of $a\omega$. In section \ref{sec:kerrae} we show
that this simple technique yields remarkably accurate results, even
for large values of $|a\omega|$.

\subsection{Results}

\begin{figure*}
\begin{center}
\begin{tabular}{cc}
\epsfig{file=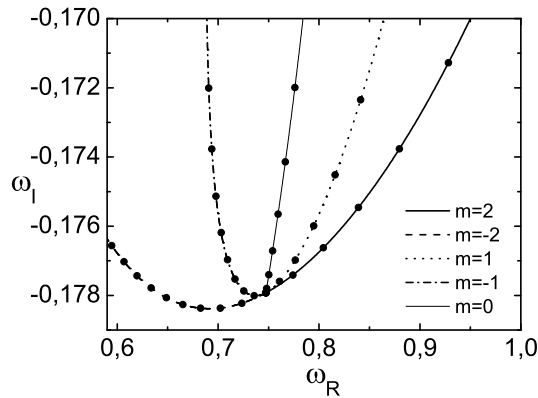,width=8cm,angle=0}
\end{tabular}
\caption{``Zeeman--like'' splitting of the fundamental gravitational
mode with $l=2$. We mark by dots the points corresponding to
$a=0,~0.05,0.10,\dots 0.5$.
\label{fig:fig5ono}}
\end{center}
\end{figure*}

The most relevant feature of the Kerr quasinormal spectrum is that
rotation acts very much like an external magnetic field on the energy
levels of an atom, causing a {\it Zeeman splitting} of the quasinormal
mode spectrum. We illustrate the splitting of the fundamental
quasinormal mode in Figure \ref{fig:fig5ono} (see also \cite{O}). As
the rotation parameter $a$ increases the branches with $m=2$ and
$m=-2$ move in opposite directions, being tangent to the the branches
with $m=1$ and $m=-1$. In this simple case the effect of rotation on
the modes is roughly proportional to $m$, as physical intuition would
suggest. The mode with $m=0$ has a different behavior: it ``moves''
very quickly along a direction in the complex plane which is roughly
perpendicular to the two branches with $m\neq 0$.

\begin{figure*}
\begin{center}
\begin{tabular}{cc}
\epsfig{file=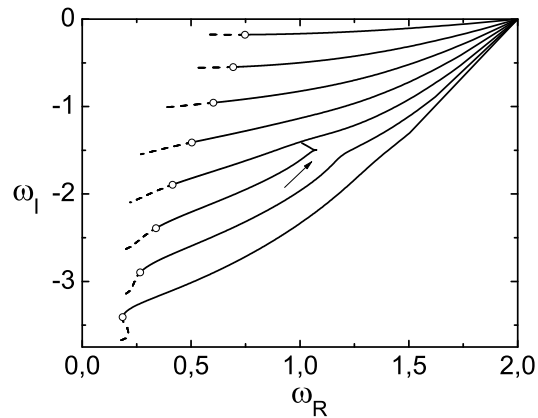,width=8cm,angle=0}
\end{tabular}
\caption{Trajectory in the complex plane of the first eight Kerr
quasinormal frequencies with $m=2$ (solid lines) and $m=-2$ (dashed
lines). Filled circles mark the corresponding mode in the
Schwarzschild limit. An arrow indicates the small loop described by
the ``exceptional'' quasinormal mode with $n=6$, that does not tend to
the critical frequency for superradiance (see also Figures 3 and 4 in
\cite{O}).
\label{fig:fig4ono}}
\end{center}
\end{figure*}

In Figure \ref{fig:fig4ono} we show the first eight quasinormal
frequencies with $m=2$ (solid lines) and $m=-2$ (dashed lines). For
large enough damping, modes with $m<0$ are not tangent to the
corresponding modes with positive $m$. Notice also that all modes with
$n<9$ and $m>0$ cluster at the critical frequency for superradiance
$\omega_{SR}=m \Omega$, where $\Omega$ is the angular velocity of the
black hole horizon. This fact was first observed by Detweiler
\cite{De}, who went on to suggest that all modes with positive $m$
should tend to the critical frequency for superradiance as the black
hole becomes extremal. It was even suggested that this could point to
a marginal instability of extremal Kerr black holes. The study of the
modes' {\it excitation amplitudes} in the same limit led to the
conclusion that such an instability does not actually occur
\cite{fm,ga2}. The mode with $n=6$ (marked by an arrow) shows a
peculiar deviation from the general trend, clearly illustrating the
fact that some positive-$m$ modes do {\it not} tend to the critical
frequency for superradiance in the extremal limit.

For gravitational wave emission, the dominant quasinormal mode is
almost certainly the fundamental mode with $l=m=2$. The reason is that
a black hole formed in a stellar collapse or in a merger event is
likely to have a rotating ``bar'' shape corresponding to spheroidal
harmonic indices $l=m=2$, and also that this mode is the most slowly
damped, as we can see from Figure \ref{fig:fig4ono}. Echeverria
\cite{eche} found that the following analytic fit reproduces numerical
data for the fundamental mode with $l=m=2$ within about $10~\%$:
\be
M\omega_R\simeq \left[1-0.63(1-a)^{3/10}\right]
\sim(0.37+0.19a)\,,
\ee
\be
\tau/M\sim\frac{4}{(1-a)^{9/10}}\left[1-0.63(1-a)^{3/10}\right]
\sim(1.48+2.09a)\,.
\ee
In these relations we restored the black hole masses; the last step is
a series expansion for small values of the rotational parameter $a$.
Fryer, Holz and Hughes produced a similar fit for the mode with $l=2$,
$m=0$ \cite{fhh}.

\begin{figure*}
\begin{center}
\begin{tabular}{cc}
\epsfig{file=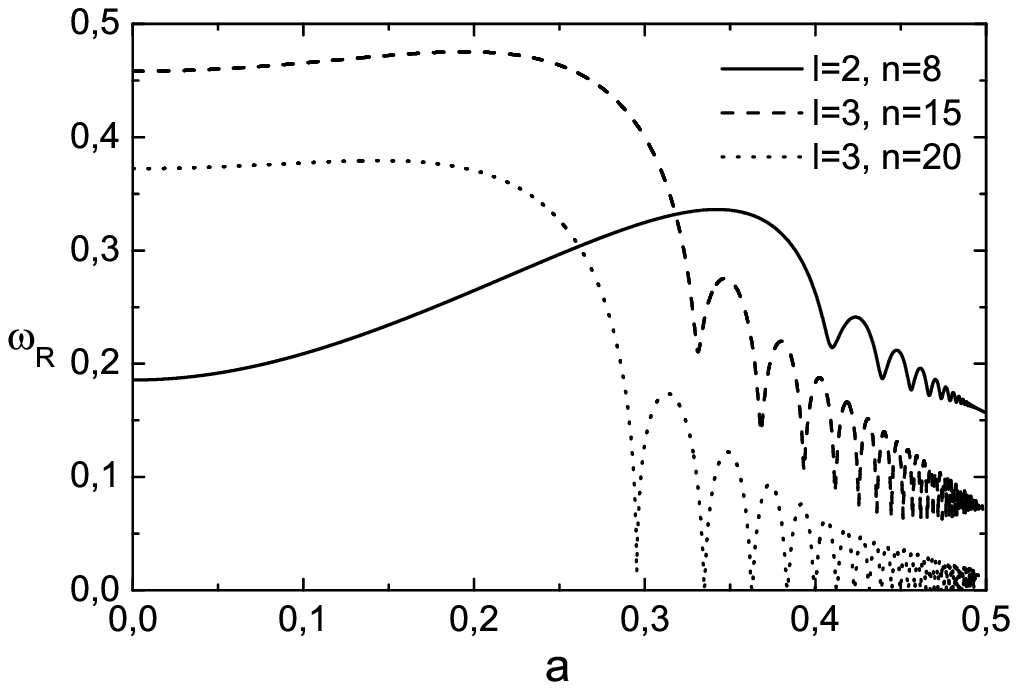,width=8cm,angle=0} &
\epsfig{file=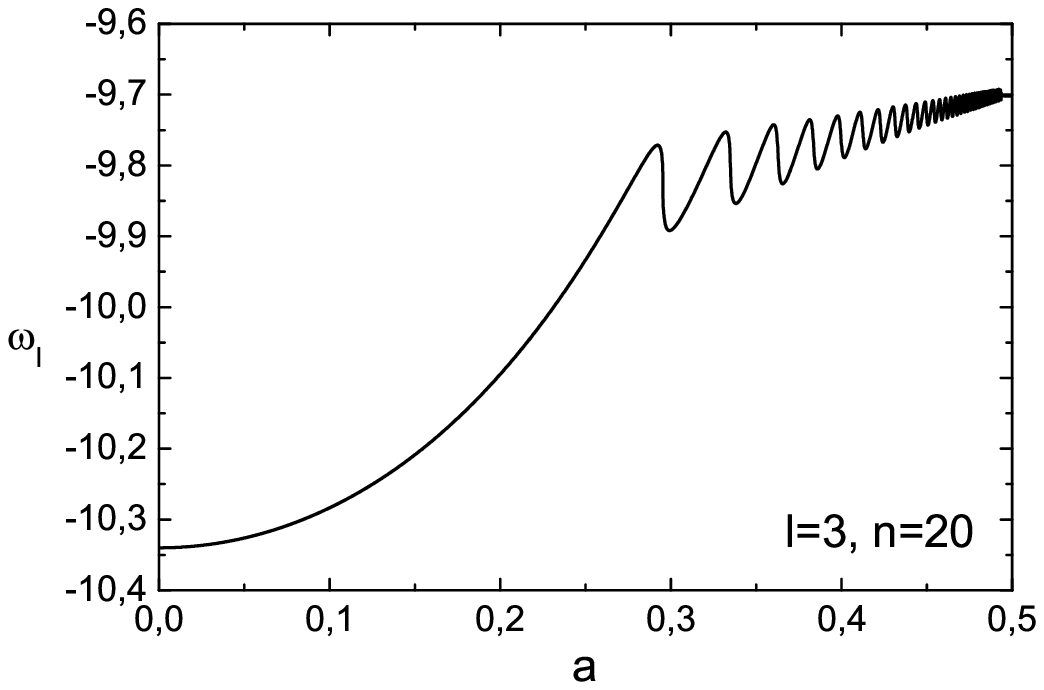,width=8cm,angle=0}
\end{tabular}
\caption{Left panel: real part of Kerr modes having $m=0$ as a
function of $a$.  Labels indicate the corresponding values of $l$ and
of the mode order $n$. Right panel: imaginary part of Kerr modes with
$l=3,~m=0,~n=20$ as a function of $a$.
\label{fig:fig5}}
\end{center}
\end{figure*}

In Figure \ref{fig:fig5} we show the real and imaginary parts of some
gravitational quasinormal frequencies with $m=0$. The looping behavior
is similar to that of RN modes modes, and the number of loops
increases with the damping of the mode. This same looping behavior was
recently found by Glampedakis and Andersson for {\it scalar}
perturbations of Kerr black holes, using a different method \cite{ga}.

\begin{figure*}
\begin{center}
\begin{tabular}{cc}
\epsfig{file=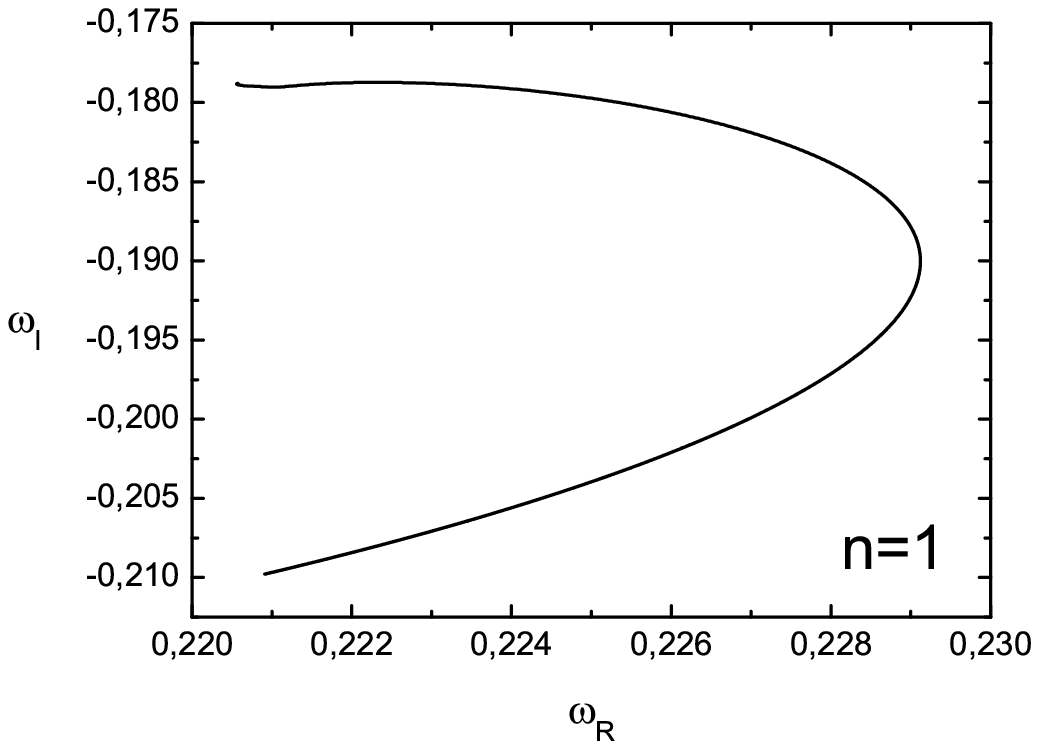,width=8cm,angle=0} &
\epsfig{file=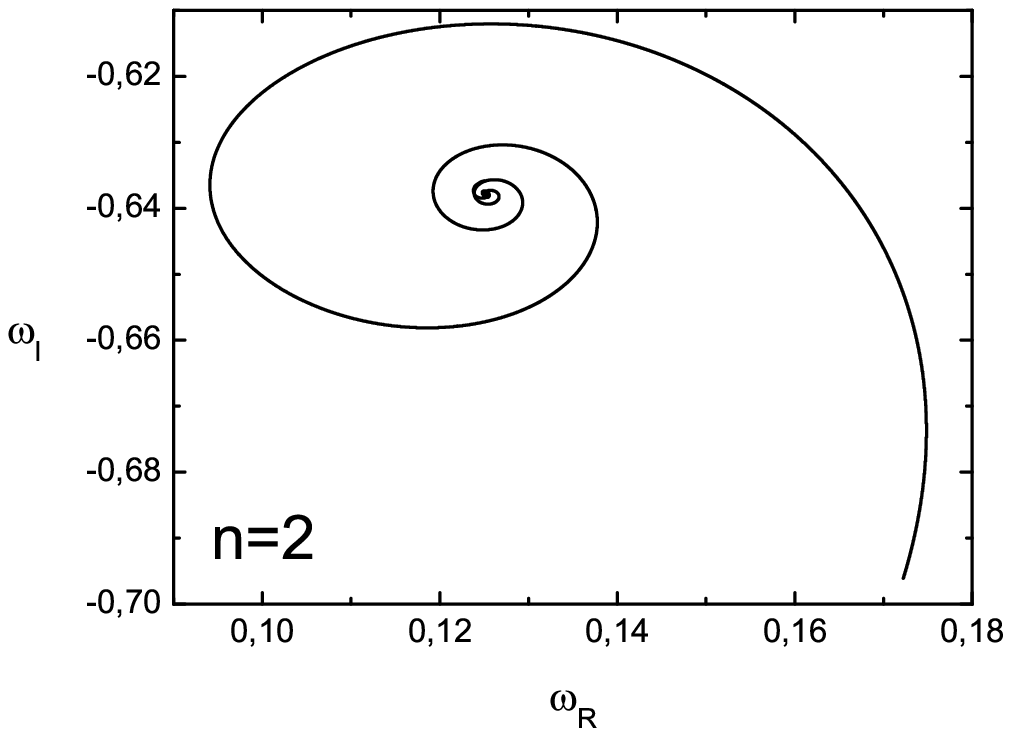,width=8cm,angle=0} \\
\epsfig{file=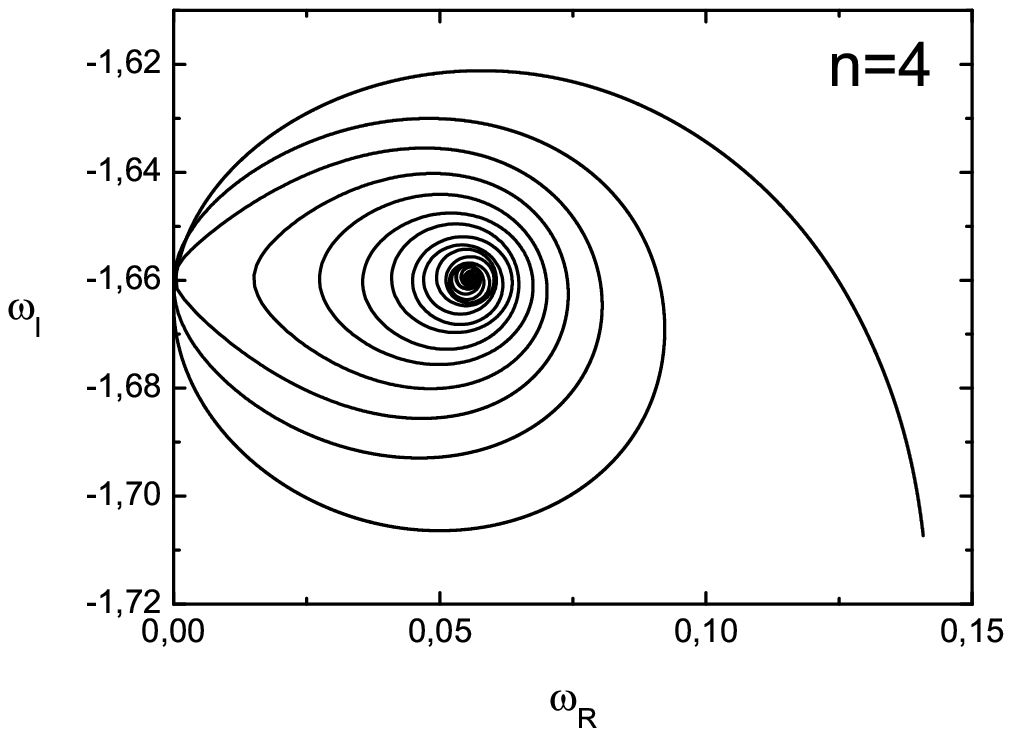,width=8cm,angle=0} &
\epsfig{file=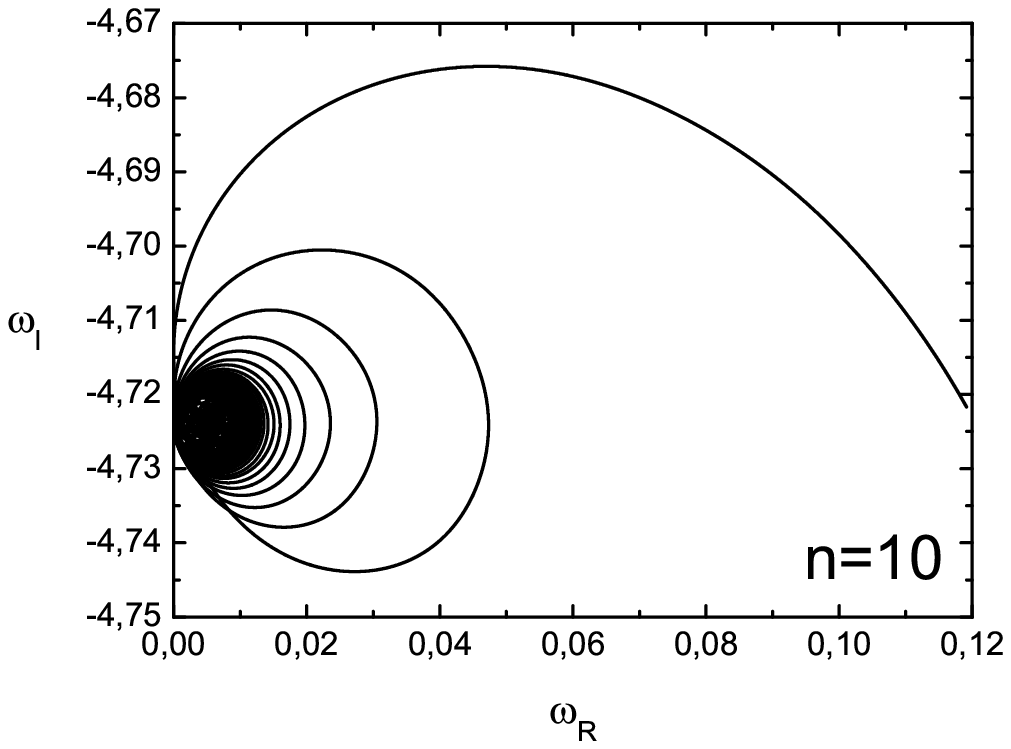,width=8cm,angle=0}
\end{tabular}
\caption{Trajectories of a few scalar modes with $l=m=0$. The
different panels correspond to the fundamental mode (top left), which
does not show a spiralling behavior, and to modes with overtone
indices $n=2,~4,~10$.
\label{fig:fig6}}
\end{center}
\end{figure*}

Figure \ref{fig:fig6} shows the trajectory in the complex plane of
some scalar modes with $m=0$. The spirals wind up very quickly as the
damping grows, and the ``center'' of the spiral rapidly moves towards
the pure-imaginary axis. In section \ref{sec:kerrhd} we will see that
the oscillation frequency of modes with $m=0$ probably tends to {\it
zero} as $|\omega_I|\to \infty$.

\begin{figure*}
\begin{center}
\begin{tabular}{cc}
\epsfig{file=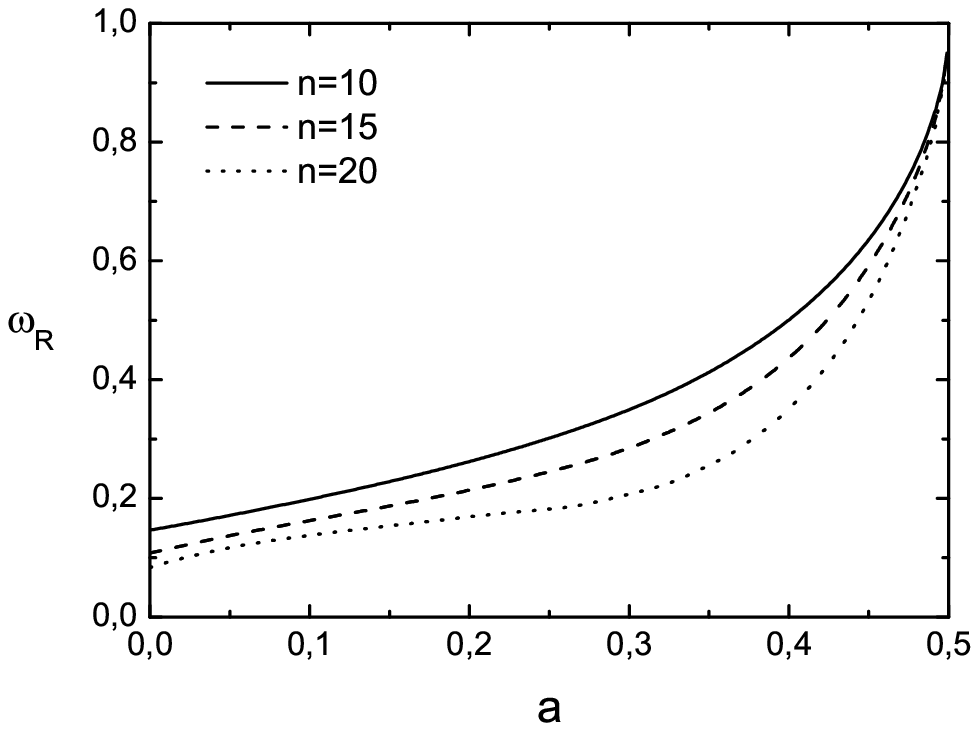,width=8cm,angle=0} &
\epsfig{file=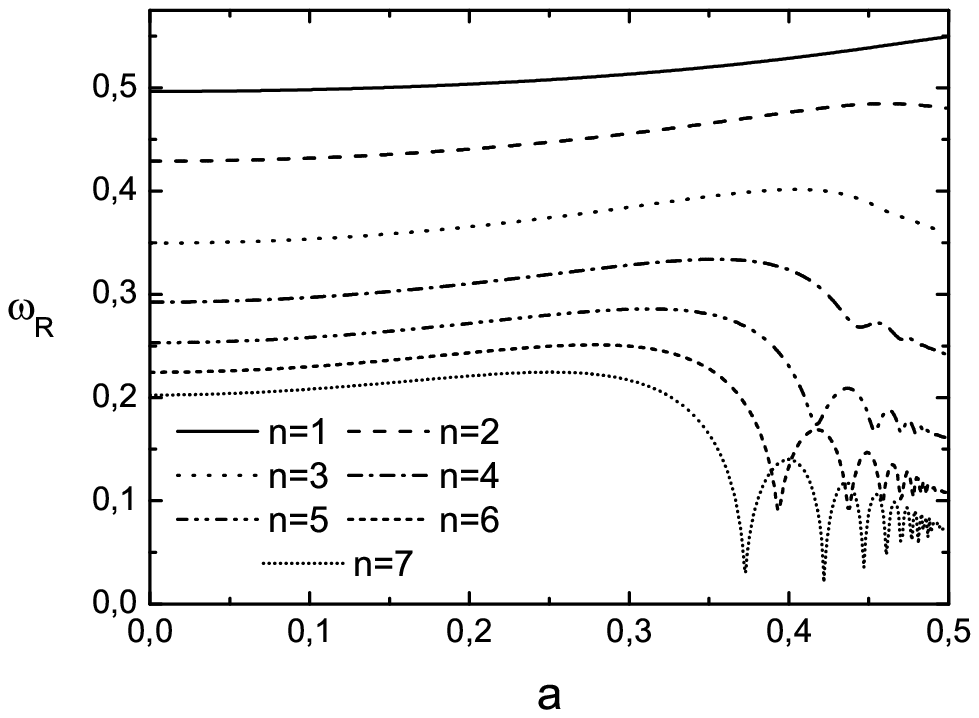,width=8cm,angle=0} \\
\epsfig{file=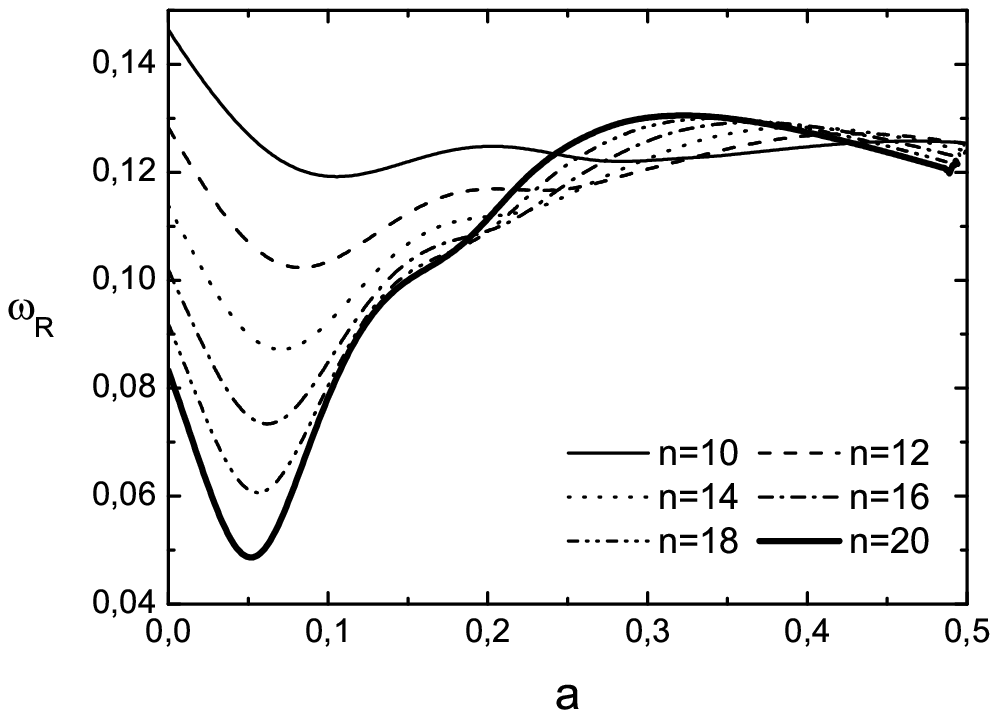,width=8cm,angle=0} &
\epsfig{file=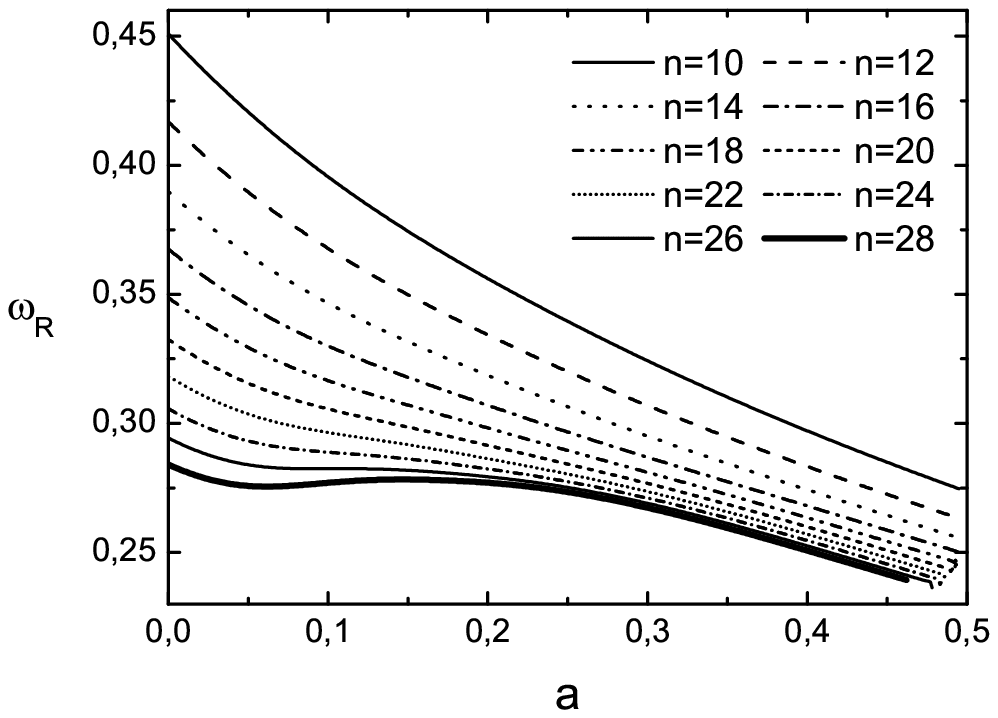,width=8cm,angle=0}
\end{tabular}
\caption{Real part of electromagnetic modes with $l=m=1$ (top left),
$l=1$, $m=0$ (top right), $l=1$, $m=-1$ (bottom left) and $l=2$,
$m=-2$ (bottom right) as a function of the rotation parameter $a$, for
increasing values of the mode index.
\label{fig:fig7}}
\end{center}
\end{figure*}

In \cite{BCKO} we found that for intermediate damping (say $10\lesssim
n \lesssim 50$) quasinormal frequencies belong to three different
families depending on whether $m>0$, $m=0$ or $m<0$. Figure
\ref{fig:fig7} shows some representative results for electromagnetic
modes ($s=-1$), but the qualitative behavior of gravitational and
scalar modes is the same:

\begin{itemize}

\item[1)] Modes with $m>0$. For $10\lesssim n \lesssim 50$, $\omega_R$
has a relative minimum as a function of $a$, but the numerics do not
show any convergence to an asymptotic limit.  Only the real part of
gravitational modes with $l=m=2$ tends to the limit $\omega_R=2
\Omega$ when $n\sim 50$ (see \cite{BK} and Figure 1 in
\cite{BCKO}). This deceiving convergence seems related to the peculiar
behavior of the angular separation constant for $|s|=l=m=2$ (see
Figure \ref{fig:fig10}). For $m>0$ most modes in the range $10\lesssim
n \lesssim 50$ approach the limiting value $\omega_R=m \Omega$ as
$a\to 1/2$, but there are exceptions (see Figure
\ref{fig:fig4ono}). The imaginary parts of all modes with $m>0$ have a
separation $2\pi T_H$, where now $T_H$ is the Hawking temperature of a
Kerr black hole. The agreement with the separation of consecutive
Schwarzschild quasinormal modes does not seem to be a coincidence,
$T_H$ being a non-trivial function of the rotation rate.

\item[2)] Modes with $m=0$. The real and imaginary parts of these
modes oscillate as a function of $a$, a behavior reminiscent of RN
quasinormal modes. As $n$ increases, $\omega_R$ becomes smaller. The
spacing between the imaginary parts oscillates and is {\it not} given
by $2\pi T_H$.

\item[3)] Modes with $m<0$. At least for $a\gtrsim 0.25$, $\omega_R$
seems to converge to a (weakly $a$--dependent) limit $\omega_R\simeq
m\varpi$, where $\varpi\sim 0.12$, whatever the value of $l$ and the
spin of the perturbing field. The spacing in the imaginary parts for
these values of $n$ does not show convergence, and is {\it not} given
by $2\pi T_H$.

\end{itemize}

These results are even more puzzling than in the RN case. To probe
the asymptotic regime we clearly need to push the calculation to
higher damping. The main problem here is that Leaver's approach
requires the {\it simultaneous} solution of the radial and angular
continued fraction conditions. For mode order $n\gtrsim 50$ the method
becomes unreliable, or it just fails to provide results. A way around
this ``coupling problem'' is to study the asymptotic behavior of the
angular equation as $|a\omega|\simeq |a\omega_I|\to \infty$. In the
next section we carry out this task. We first compute numerically
$_sA_{lm}(a\omega)$ when $\omega\simeq \ii \omega_I$ and $|a\omega|\to
\infty$. Then we determine analytically the leading-order behavior of
the separation constant. Finally, in section \ref{sec:kerrhd} we
replace this analytic expansion in the radial continued fraction,
effectively getting rid of the angular equation. By this trick we can
find quasinormal frequencies with $n\gg 50$ solving {\it only} the
radial continued fraction, and ultimately determine the asymptotic
behavior of $\omega_R(a)$ for Kerr black holes.

%%%%%%%%%%%%%%%%%%%%%%%%%%%%%%%%%%%%%%%%%%%%%%%%%%%%%%%%%%%%%%%%%%%%%%%%%%%%%%%

\subsection{Angular eigenvalues}
\label{sec:kerrae}

The analytical properties of the angular equation
(\ref{angularwaveeq}) and of its eigenvalues have been studied by many
authors
\cite{flammer,li,barrowes,falloon,early,seidel,breuerbook,breuer}.
Series expansions of $_{s}A_{lm}$ for $|a \omega|\ll 1$ have long been
available, and they agree well with numerical results (see
eg. \cite{seidel}, where some mistakes and notational differences in
the previous literature are discussed and corrected). 

Series expansions for large $|a \omega|$ have been known for a long
time in the special case of $s=0$. Such an analytical power-series
expansion for large (pure real and pure imaginary) values of $a
\omega$ can be found, for example, in Flammer's book
\cite{flammer}. Flammer's results are in good agreement with the
pioneering numerical work by Oguchi \cite{oguchi}, who computed
angular eigenvalues for complex values of $a \omega$ and $s=0$. A
review of numerical methods to compute eigenvalues and eigenfunctions
for $s=0$ can be found in \cite{li}.  Recently Barrowes {\it et al.}
studied the large frequency behavior of $s=0$ harmonics in the complex
frequency plane \cite{barrowes}, while Falloon {\it et al.} developed
a {\it Mathematica} notebook to compute $s=0$ harmonics for general
complex values of the frequency \cite{falloon}.

Unfortunately similar results for large values of $|a \omega|$ and
general spin $s$ are lacking. Even worse, until recently the few
analytical predictions were contradictory \cite{comment}. The
situation for large {\it real} frequencies has finally been clarified
by Casals and Ottewill \cite{casals}, who corrected some mistakes in
reference \cite{breuer}. Here we fill the remaining gap. We present
numerical and analytical results for $_{s}A_{lm}$ when the frequency
$a \omega$ is large and {\it purely imaginary}. In Flammer's language
we are looking at the {\it prolate} case, while Casals and Ottewill
deal with the {\it oblate} case.

\begin{figure*}
\begin{center}
\begin{tabular}{cc}
\epsfig{file=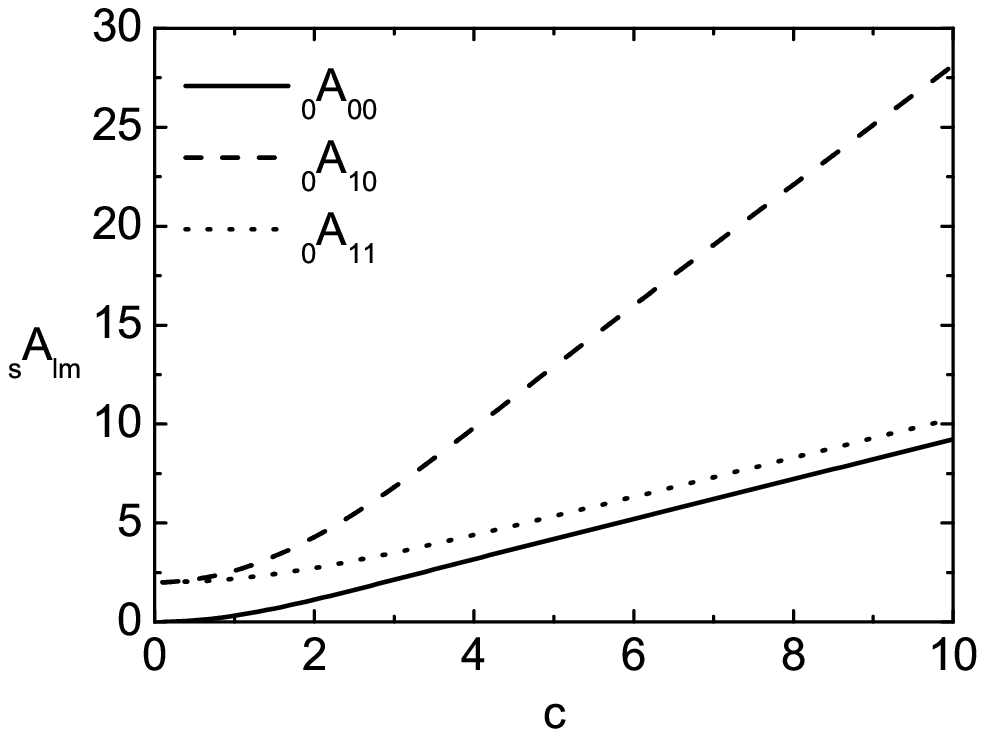,width=8cm,angle=0} &
\epsfig{file=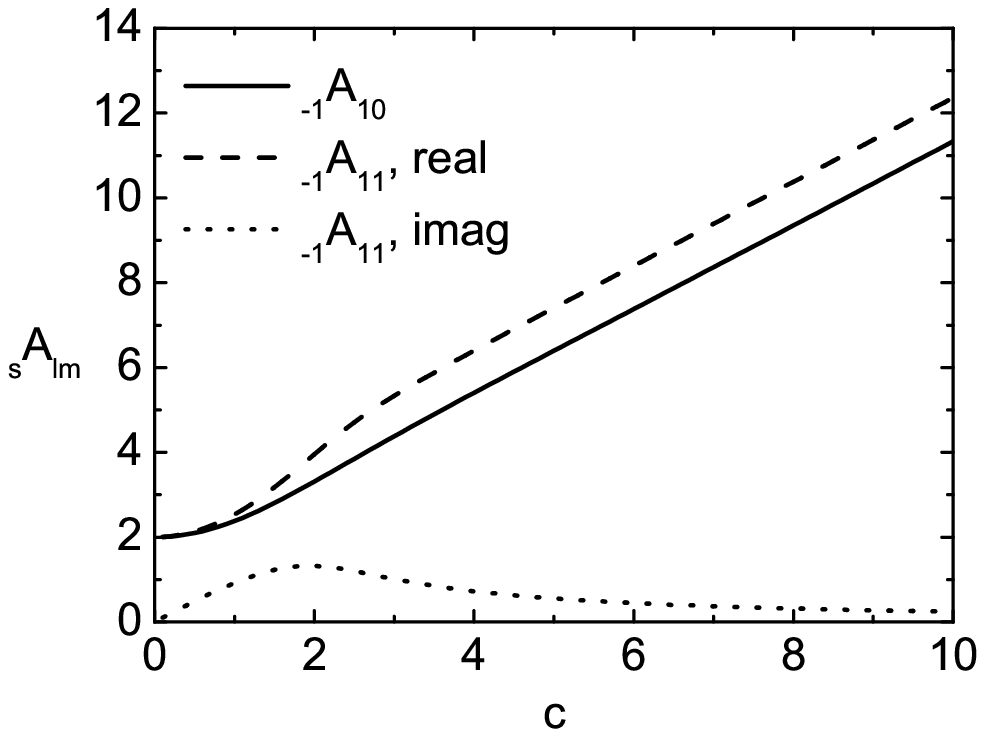,width=8cm,angle=0} \\
\epsfig{file=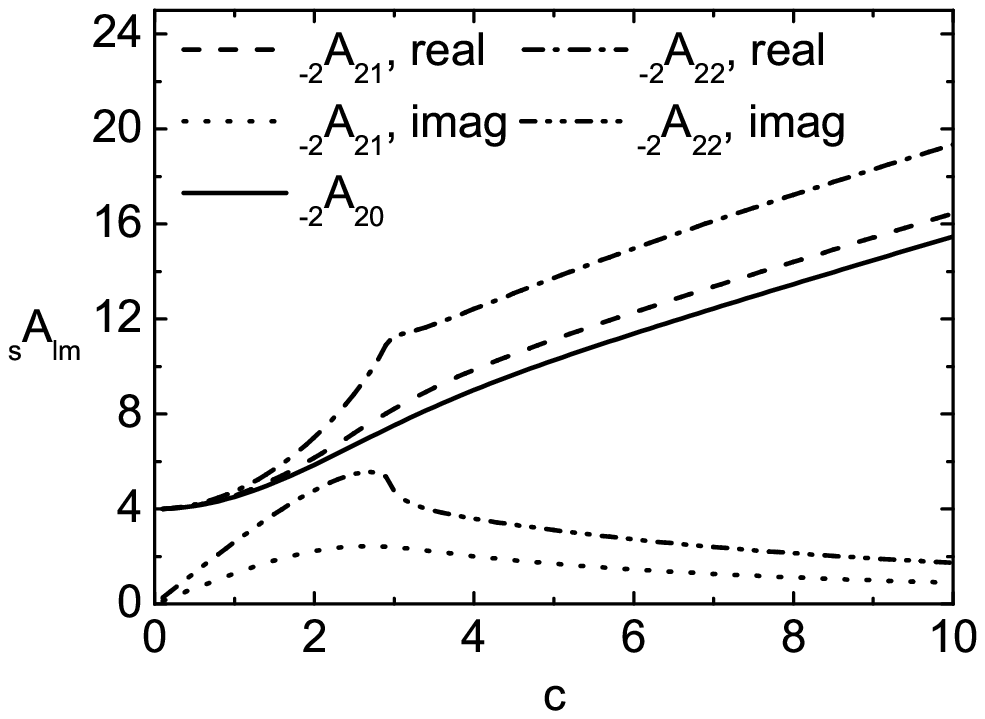,width=8cm,angle=0} &
\epsfig{file=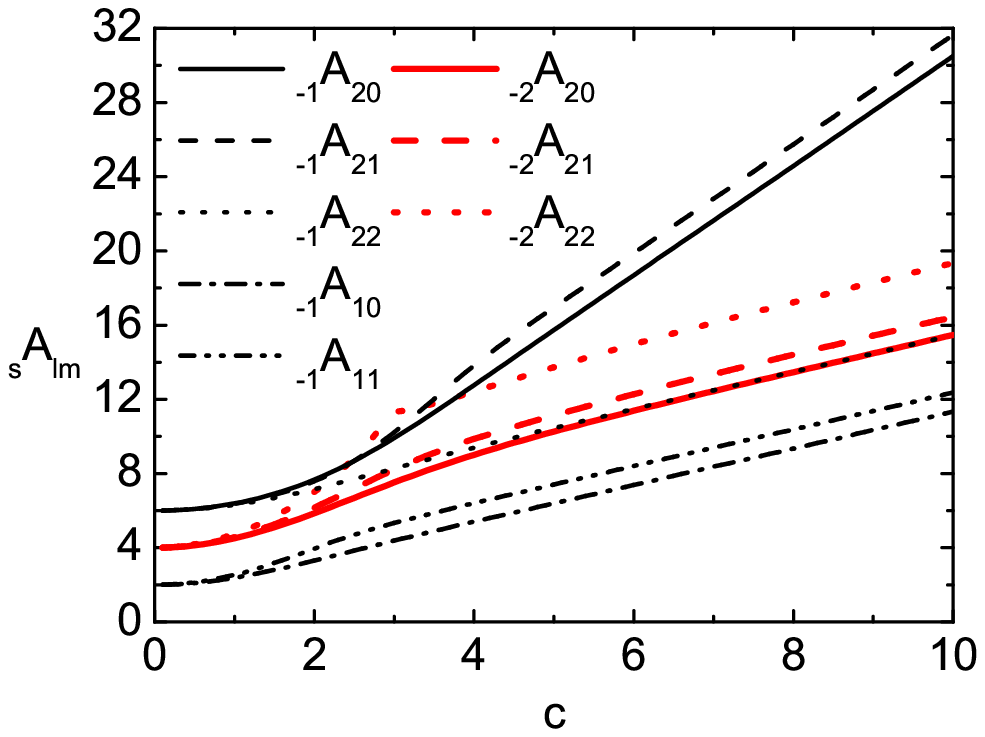,width=8cm,angle=0}
\end{tabular}
\caption{Top left: angular separation constant $_sA_{lm}$ for $s=0$ as
a function of $c=a\omega_I$. The numerical behavior is in agreement
with Flammer's asymptotic formula for the prolate case, and the
separation constant is purely real. The slopes $k_{lm}$ in
$_sA_{lm}\sim k_{lm} c$ are either 1 or 3. Top right, bottom left:
angular separation constant for $s=-1$ and $s=-2$. Now $_sA_{lm}$ is
not purely real, but because of eq. (\ref{conj}) it is sufficient to
compute eigenvalues with $m>0$. In both plots, lines starting from
$_sA_{lm}=l(l+1)-s(s+1)$ are the real parts, and lines starting from
$_sA_{lm}=0$ are the imaginary parts. Notice the ``kink'' in the real
part at $c \simeq 3$ when $s=-2$ and $l=m=2$: there is no
discontinuity, just a change of slope.  This peculiar behavior might
explain the deceiving convergence of the quasinormal frequency to
$\omega_R=2\Omega$ that we observed in \cite{BK} for $l=m=2$ and
$s=-2$. Bottom right: real parts of $_sA_{lm}$ for $s=-1$ (in black)
and $s=-2$ (in red). The slopes for the linear growth at large $c$ are
all consistent with Equation (\ref{alm}).
\label{fig:fig10}}
\end{center}
\end{figure*}

We carry out our numerical calculations choosing $a\omega$ to be pure
imaginary and solving Leaver's angular continued fraction as explained
in section \ref{kcm}. Our numerical values for $s=0$ and purely
imaginary frequency (top left panel in Figure \ref{fig:fig10}) are in
agreement with Tables 10-12 in \cite{flammer}. For complex $\omega$
and $s=0$ we could reproduce Table 2 in \cite{oguchi}. As a final
sanity check, we verified in a few representative cases that the {\it
Mathematica} notebooks presented in \cite{li, falloon} are in
agreement with our code when we specialize to $s=0$.

We also verified numerically some elementary analytical properties of
the eigenvalues: for example, the computed eigenvalues satisfy the
relation $_{-s}A_{lm}=_{s}A_{lm}+2s$. In the oblate case considered in
\cite{casals} the eigenvalues grow quadratically, $_{s}A_{lm}\sim
-|a\omega|^2$, as they should \cite{casals}. Preliminary comparisons
with Casals show excellent agreement on the computed eigenvalues for
all values of $s$ \cite{bcc}.

For $s=0$ the series expansion given by formula (8.1.11) in
\cite{flammer} is an excellent approximation to the numerical data:
\begin{eqnarray}
_{0}A_{lm}&=&(2L+1)c-(2L^2+2L+3-4m^2)2^{-2}
-(2L+1)(L^2+L+3-8m^2)2^{-4}c^{-1} \\
&-&[5(L^4+2L^3+7L+3)-48m^2(2L^2+2L+1)]2^{-6}c^{-2}+{} + {\cal
O}(c^{-3})\,.\nonumber
\end{eqnarray}
Here we defined $c=a\omega_I$ and $L=l-|m|$. Comparison with numerical
results (see the top left panel of Figure \ref{fig:fig10}) shows that
this expansion is reasonably accurate down to $c\simeq 2$.

For $s=-1$ and $s=-2$ the angular separation constant is complex. We
can limit our calculations to positive $m$'s because of the following
symmetry property (see eg. \cite{L}):
\begin{equation}\label{conj}
_sA_{l~-m}= _sA_{lm}^*\,,
\end{equation}
where the asterisk denotes complex conjugation. Numerical results for
the lowest radiatable multipoles are given in Figure
\ref{fig:fig10}. There we plot together real and imaginary parts of
the eigenvalues. It is apparent that the real part is dominant,
growing linearly as $c\to \infty$.

The proportionality constant for this linear growth can be found by a
straightforward generalization of Flammer's method to general values
of $s\neq 0$. Define a new angular wavefunction $Z_{lm}(u)$ through
\cite{breuerbook}
\begin{equation} 
S_{lm}(u)=(1-u^2)^{\frac{m+s}{2}}Z_{lm}(u)\,,
\label{newwave}
\end{equation} 
and change independent variable by defining $x=\sqrt{2c}{u}$, where
$c^2=-(a\omega)^2$.  Substitute this in (\ref{angularwaveeq}) to get:
\begin{eqnarray} 
\left[ _{s}A_{lm}-\frac{cx^2}{2}-i\sqrt{2c}x-m(m+1)-\frac{2msx}{\sqrt{2c}+x}\right] Z_{lm}+
(2c-x^2)Z_{lm,xx}-2(m+s+1)x Z_{lm,x}=0\,.
\label{angwaveeq2}
\end{eqnarray}
When $c \rightarrow \infty$, this equation becomes a parabolic
cylinder function.  The arguments presented in
\cite{flammer,breuerbook,breuer} lead to
\begin{equation}
_{s}A_{lm}=(2L+1)c+ {\cal O}(c^0) \,\,,\, c \rightarrow \infty \,,
\label{alm}
\end{equation}
where $L$ is the number of zeros of the angular wavefunction inside
the domain.  One can show that
\begin{equation}
L=\left\{ \begin{array}{ll}
            l-|m|\,,   & {|m| \geq |s|},\\ 
            l-|s|\,,   &{|m|<|s|}. 
\end{array}\right.
\label{jdef}
\end{equation}
Higher order corrections in the asymptotic expansion can be obtained
as indicated in \cite{flammer}. We will not need them in our
calculation of highly damped modes.

%%%%%%%%%%%%%%%%%%%%%%%%%%%%%%%%%%%%%%%%%%%%%%%%%%%%%%%%%%%%%%%%%%%%%%%%%%%%%%%

\subsection{High damping}
\label{sec:kerrhd}

To compute the asymptotic quasinormal frequencies of the Kerr black
hole we use a technique similar to that described in \cite{N}. We fix
a value of the rotation parameter $a$. We first compute quasinormal
frequencies for which $|a \omega|\sim 1$, so that formula (\ref{alm})
is only marginally valid. This procedure is consistent with our
previous intermediate-damping calculations: for example, when we
include terms up to order $|a\omega|^{-2}$ in the asymptotic expansion
for $_{0}A_{lm}$ provided in \cite{flammer}, our new results for
$a\simeq 0.1$ and $l=m=2$ match the results for the scalar case at
overtone numbers $20\lesssim n\lesssim 30$.  Then we increase the
overtone index $n$ (progressively increasing the inversion index of
our ``decoupled'' continued fraction). Finally, we fit our numerical
results by functional relations of the form:
\begin{equation}
\omega_R(n)=\omega_R+\omega_R^{(1)}\alpha+\omega_R^{(2)}\alpha^2
+\omega_R^{(3)}\alpha^3,
\nonumber
\end{equation}
where $\alpha=1/|\omega_I|$ or $\alpha=\sqrt{1/|\omega_I|}$.  At
variance with the non-rotating case \cite{N}, fits in powers of
$1/|\omega_I|$ perform better, especially for small and large
$a$. However, both fits break down as $a\to 0$: the values of
higher-order fitting coefficients increase in this limit, so that
subdominant terms become as important as the leading order, and the
extraction of the asymptotic frequency $\omega_R$ becomes
problematic. The numerical behavior of subdominant coefficients
supports the expectation (which has not yet been verified
analytically) that subdominant corrections are
$a$-dependent. Therefore one has to be careful to the order in which
the limits $n\to \infty$, $a\to 0$ are taken \cite{MN,BK,BCKO}. These
observations are consistent with the fact that, in the RN case, the
zero-charge limit of the asymptotic quasinormal frequency spectrum
does not yield the asymptotic Schwarzschild quasinormal spectrum.

\begin{figure*}
\begin{center}
\begin{tabular}{cc}
\epsfig{file=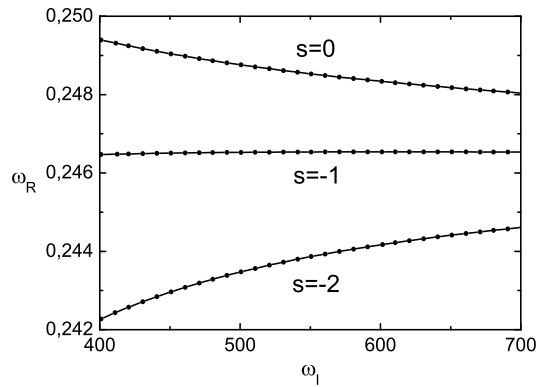,width=8cm,angle=0}
\end{tabular}
\caption{Real part of high-order quasinormal frequencies for scalar
($s=0$), electromagnetic ($s=-1$) and gravitational ($s=-2$)
perturbations of a Kerr BH with $a=0.1$ ($l=m=2$). Quasinormal
frequencies of different spins converge to the same value. For any
kind of perturbation we are already deep in the region of validity of
the asymptotic expansion (\ref{alm}), since $|a\omega| \sim 60$.
\label{fig:fig11}}
\end{center}
\end{figure*}

\begin{figure*}
\begin{center}
\begin{tabular}{cc}
\epsfig{file=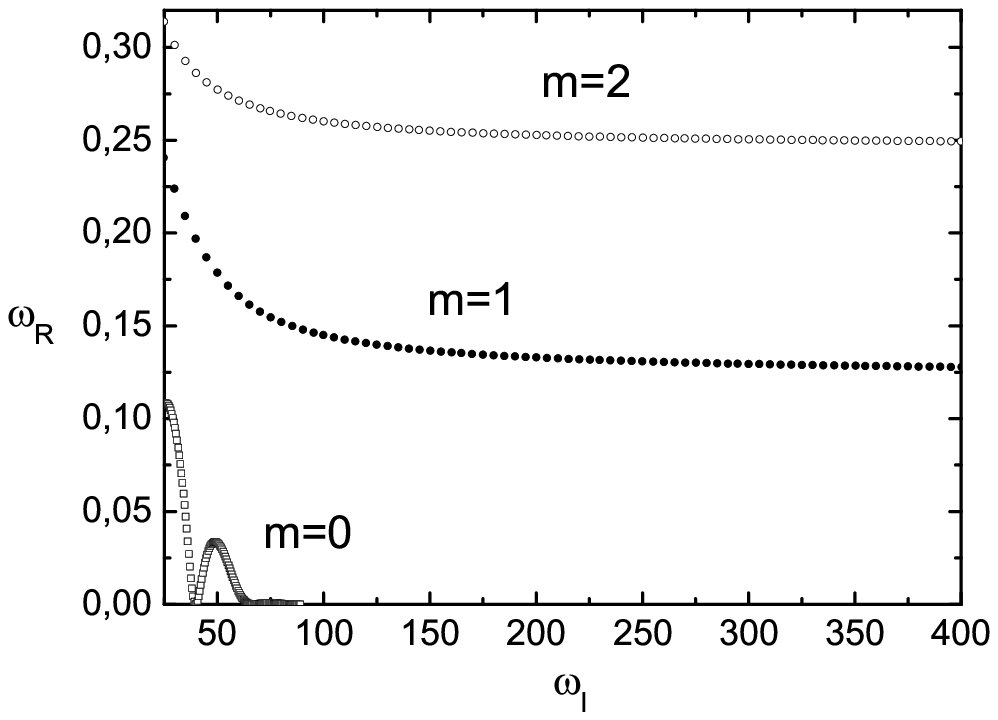,width=8cm,angle=0} &
\epsfig{file=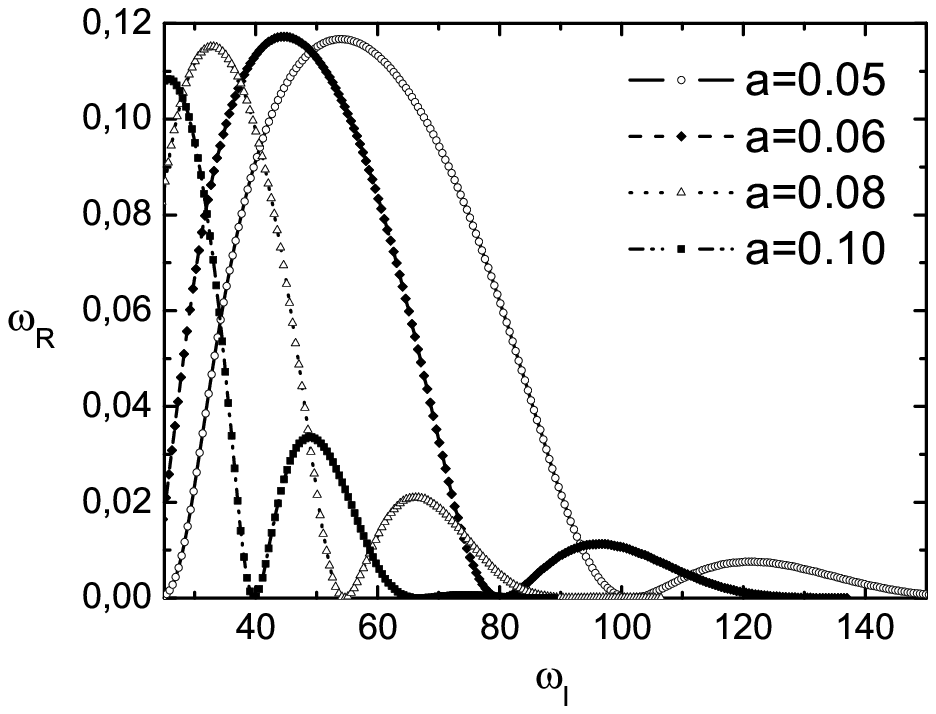,width=8cm,angle=0}
\end{tabular}
\caption{Gravitational quasinormal frequencies of a Kerr BH with
$a=0.1$, $l=2$ and different values of $m$. Dashed lines and open
circles correspond to negative $m$. The asymptotic value does not
depend on the sign of $m$, and is proportional to $|m|$.  On the right
we show $m=0$ quasinormal frequencies for different values of the
rotation parameter: the larger $a$, the faster $\omega_R$ tends to
zero as $|\omega_I|\to \infty$.
\label{fig:fig12}}
\end{center}
\end{figure*}

We have extracted asymptotic frequencies using two independent
numerical codes.  For each value of $a$, we found that the
extrapolated value of $\omega_R$ is independent of $s$
(Fig. \ref{fig:fig11}), independent of $l$ and proportional to $m$
(Fig. \ref{fig:fig12}): 
\be
\omega_R=m \varpi(a)\,.
\ee

\begin{figure*}
\begin{center}
\begin{tabular}{cc}
\epsfig{file=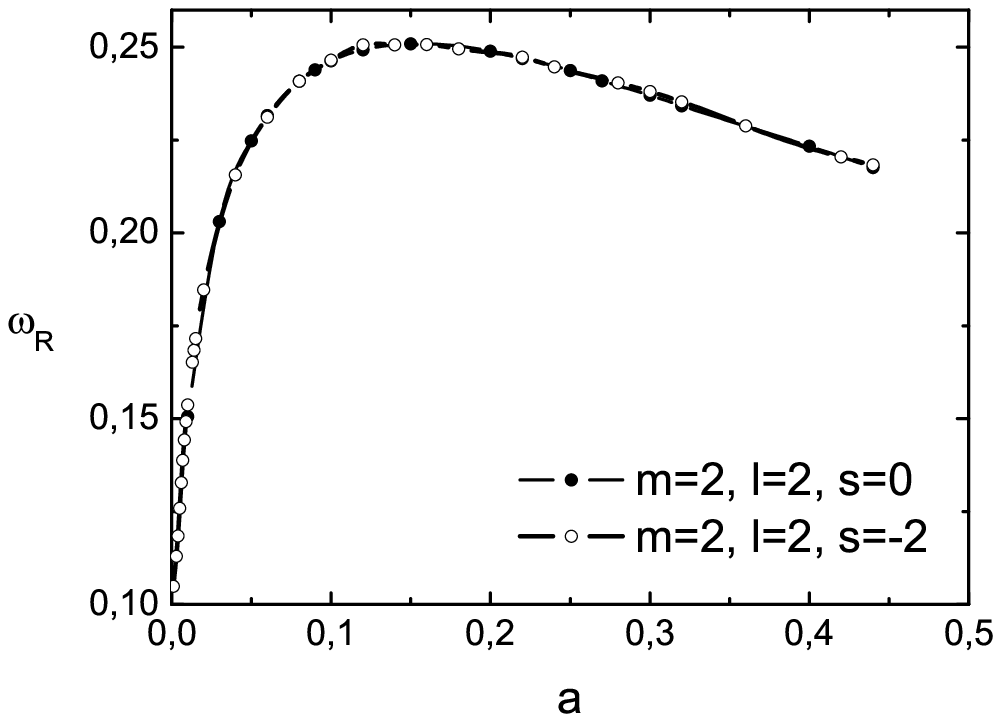,width=8cm,angle=0} &
\epsfig{file=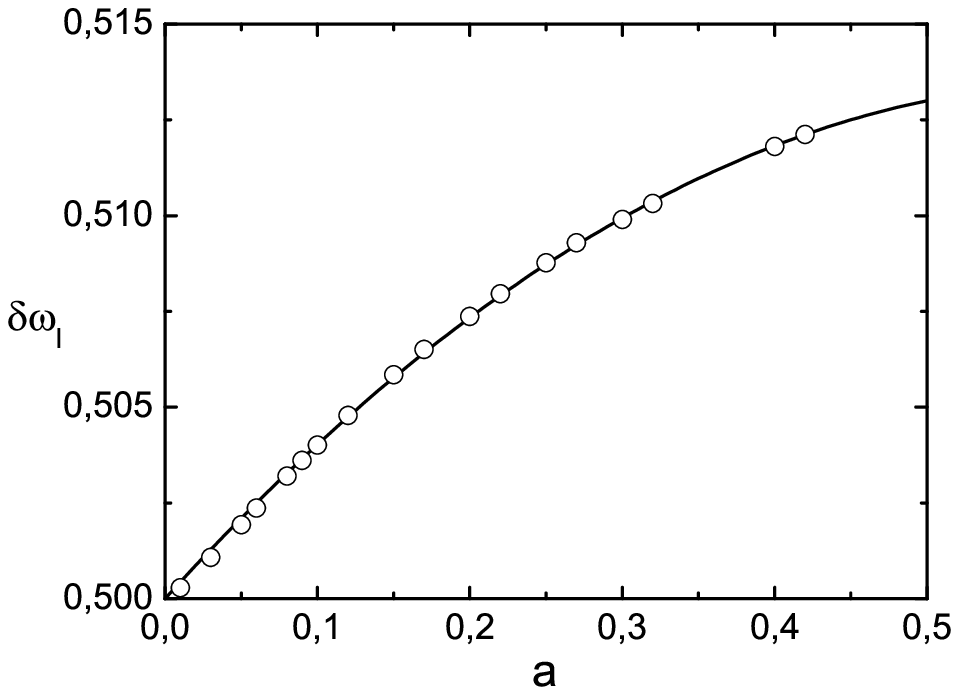,width=8cm,angle=0}
\end{tabular}
\caption{Asymptotic real part $\omega_R=2\varpi(a)$ of the $l=m=2$
gravitational and scalar quasinormal frequencies extrapolated from
numerical data: $\omega_R\to 2\varpi(1/2)\simeq 0.21$ as $a\to
1/2$. Results are independent of $l$, $s$ and the sign of $m$.
\label{fig:fig13}}
\end{center}
\end{figure*}

We obtained $\omega_R$ computing quasinormal frequencies both for
scalar perturbations ($s=0$) and for gravitational perturbations
($s=-2$). In both cases we picked $l=m=2$. The agreement between the
extrapolated behaviors of $\omega_R$ as a function of $a$ is
excellent, suggesting that both sets of results are typically reliable
with an error $\lesssim 1 \%$.  Our results are also weakly dependent
on the number of terms used in the asymptotic expansion of $_{s}A_{l
m}$: this provides another powerful consistency check. We have tried
to fit the resulting ``universal function'', displayed in
Fig. \ref{fig:fig13}, by simple polynomials in the BH's Hawking
temperature $T$ and angular velocity $\Omega$ (and their
inverses). None of these fits reproduces our numbers with satisfactory
accuracy.  It is quite likely that asymptotic quasinormal frequencies
will be given by an implicit formula involving the exponential of the
Kerr black hole temperature, as in the RN case \cite{MN,AH,BK}.

For any $a$, the imaginary part $\omega_I$ grows without bound. Quite
surprisingly, the spacing between modes $\delta \omega_I$ is a
monotonically increasing function of $a$: it is not simply given by
$2\pi T_H$, as some numerical and analytical calculations suggest
\cite{BK,BCKO,spacing}. A power fit in $a$ of our numerical results
yields:
\begin{equation}
\delta \omega_I=1/2+0.0438a-0.0356a^2. 
\end{equation}

%%%%%%%%%%%%%%%%%%%%%%%%%%%%%%%%%%%%%%%%%%%%%%%%%%%%%%%%%%%%%%%%%%%%%%%%%%%%%%%

\section{Algebraically special modes}
\label{sec:kerras}

Algebraically special modes of Schwarzschild black holes have been
studied for a long time, but only recently their understanding has
reached a satisfactory level. Among the early studies rank those of
Wald \cite{W} and of Chandrasekhar \cite{Cas}, who gave the exact
solution of the Regge--Wheeler, Zerilli and Teukolsky equations at the
algebraically special frequency. The nature of the quasinormal mode
boundary conditions at the Schwarzschild algebraically special
frequency is extremely subtle, and has been studied in detail by
Maassen van den Brink \cite{MVDBas}. 

To understand the problem, it is useful to recall that black hole
oscillation modes can be classified into three groups (here we use
Maassen van den Brink's ``observer-centered definition'' of the
boundary conditions):

1) ``standard'' quasinormal modes, which have outgoing wave boundary
conditions at both sides (that is, they are outgoing at infinity and
``outgoing into the horizon'');

2) total transmission modes from the left (TTM$_L$'s) are incoming
from the left (the black hole horizon) and outgoing to the other side
(spatial infinity);

3) total transmission modes from the right (TTM$_R$'s) are incoming
from the right and outgoing to the other side.

The Schwarzschild ``algebraically special'' frequency is given by
formula (\ref{AlgSp}) for Schwarzschild black holes [or by formula
(\ref{RNas}) for RN black holes], and traditionally it has been
associated with TTM's. In fact, the algebraically special frequency is
so called precisely because of the requirement that the perturbations
should be ``special'': they should contain only ingoing waves (the
perturbed Weyl scalar $\Psi_0\neq 0$) or only outgoing waves
($\Psi_4\neq 0$). Speciality is equivalent to the condition that a
certain quantity (the Starobinsky constant) should be zero. When
Chandrasekhar determined the perturbation satisfying this condition
\cite{Cas} he did not check that it does indeed satisfy TTM boundary
conditions. It turns out that, in general, it does not
\cite{MVDBas}. The Regge--Wheeler equation and the Zerilli equation
(which are known to yield the same quasinormal mode spectrum, being
related by a supersymmetry transformation) have to be treated on
different footing at $\tilde \Omega_l$, since the supersymmetry
transformation leading to the proof of isospectrality is singular
there. In particular, the Regge-Wheeler equation has {\it no modes at
all} at $\tilde \Omega_l$, while the Zerilli equation has {\it both a
quasinormal mode and a TTM$_L$}.

Numerical calculations of algebraically special modes have yielded
some puzzling results. Studying the Regge-Wheeler equation (that
should have no quasinormal modes at all according to Maassen van den
Brink's analysis) and not the Zerilli equation, Leaver \cite{L} found
a quasinormal mode which is very close, but not exactly located {\it
at}, the algebraically special frequency (Table \ref{tab1}). Namely,
he found quasinormal modes at frequencies $\tilde \Omega'_l$ such that
\be
\tilde \Omega'_2=0.000000-3.998000\ii, \qquad
\tilde \Omega'_3=-0.000259-20.015653\ii.
\ee
Notice that the ``special'' quasinormal modes $\tilde \Omega'_l$ are
such that $\Re~\ii\tilde \Omega'_2<|\tilde \Omega_2|$, $\Re~\ii\tilde
\Omega'_3>|\tilde \Omega_3|$, and that the real part of $\tilde
\Omega'_3$ is not zero. Similarly, in the extremal RN case we find a
quasinormal frequency that is very close, but {\it not exactly equal
to}, the algebraically special frequency found by
Chandrasekhar. Maassen van den Brink \cite{MVDBas} speculated that the
numerical calculations may be inaccurate and that no conclusion can be
drawn on the coincidence of $\tilde \Omega_l$ and $\tilde \Omega'_l$,
{\it ``if the latter does exist at all''}.

An independent calculation was carried out by Andersson
\cite{A}. Using a phase--integral method, he found that the
Regge--Wheeler equation has pure imaginary TTM$_R$'s which are very
close to $\tilde \Omega_l$ for $2\leq l\leq 6$. He therefore suggested
that the modes he found coincide with $\tilde \Omega_l$, which would
then be a TTM.  Maassen van den Brink \cite{MVDBas} observed again
that, if all figures in the computed modes are significant, the
coincidence of TTM's and quasinormal modes is not confirmed by this
calculation, since $\tilde \Omega'_l$ and $\tilde \Omega_l$ are
numerically (slightly) different.

%He is also very critical with the WKB approximation used in
%Nils' paper (see page 12, end of right column).

Onozawa \cite{O} showed that the Kerr mode with overtone index $n=9$
tends to $\tilde \Omega_l$ as $a\to 0$, but suggested that modes
approaching $\tilde \Omega_l$ from the left and the right may cancel
each other at $a=0$, leaving only the special (TTM) mode. He also
calculated this (TTM) special mode for Kerr black holes, solving the
relevant condition that the Starobinsky constant should be zero and
finding the angular separation constant by a continued fraction
method; his results improved upon the accuracy of those in \cite{Cas}.

%Van den Brink supports Hisashi's view, and uses the P\"oschl-Teller
%potential as an example supporting the ``cancellation hypothesis''
% (page 2 of his paper, bottom of left column). However, now we know
%that the situation is somewhat more complicated than Hisashi thought
%at the time!

The analytical approach adopted in \cite{MVDBas} clarified many
aspects of the problem for Schwarzschild black holes, but the
situation concerning Kerr modes branching from the algebraically
special Schwarzchild mode is still far from clear. In \cite{MVDBas}
Maassen van den Brink, using slow--rotation expansions of the
perturbation equations, drew two basic conclusions on these modes. The
first is that, for $a>0$, the so--called Kerr special modes (that is,
solutions to the condition that the Starobinsky constant should be
zero \cite{Cas,O}) are all TTM's (left or right, depending on the sign
of $s$). The TTM$_R$'s cannot survive as $a\to 0$, since they do not
exist in the Schwarzschild limit; this is related to the limit $a\to
0$ being a very tricky one. In particular, in this limit, the special
Kerr mode becomes a TTM$_L$ for $s=-2$; furthermore, the special mode
and the TTM$_R$ cancel each other for $s=2$. Studying the limit $a\to
0$ in detail, Maassen van den Brink reached a second conclusion: the
Schwarzschild special frequency $\tilde \Omega_l$ should be a limit
point for a multiplet of ``standard'' Kerr quasinormal modes, which
for small $a$ are well approximated by
\be\label{VDBsmalla}
\omega=-4\ii-{33078176\over 700009}ma+{3492608\over 41177}\ii a^2
+{\cal O}(ma^2)
+{\cal O}(a^4)
\ee
when $l=2$, and by a more complicated formula -- his equation (7.33)
-- when $l>2$. We numerically found quasinormal modes close to the
algebraically special frequency, but they do {\it not} agree with this
analytic prediction (not even when the rotation rate $a$ is small).

\begin{figure*}
\begin{center}
\begin{tabular}{cc}
\epsfig{file=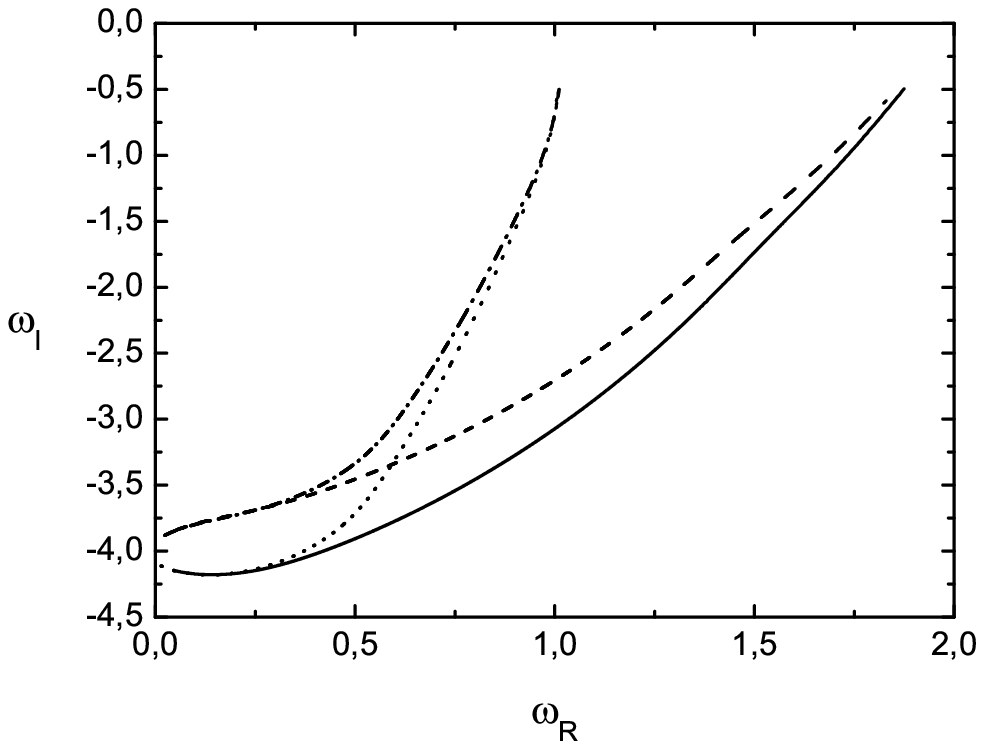,width=8cm,angle=0} &
\epsfig{file=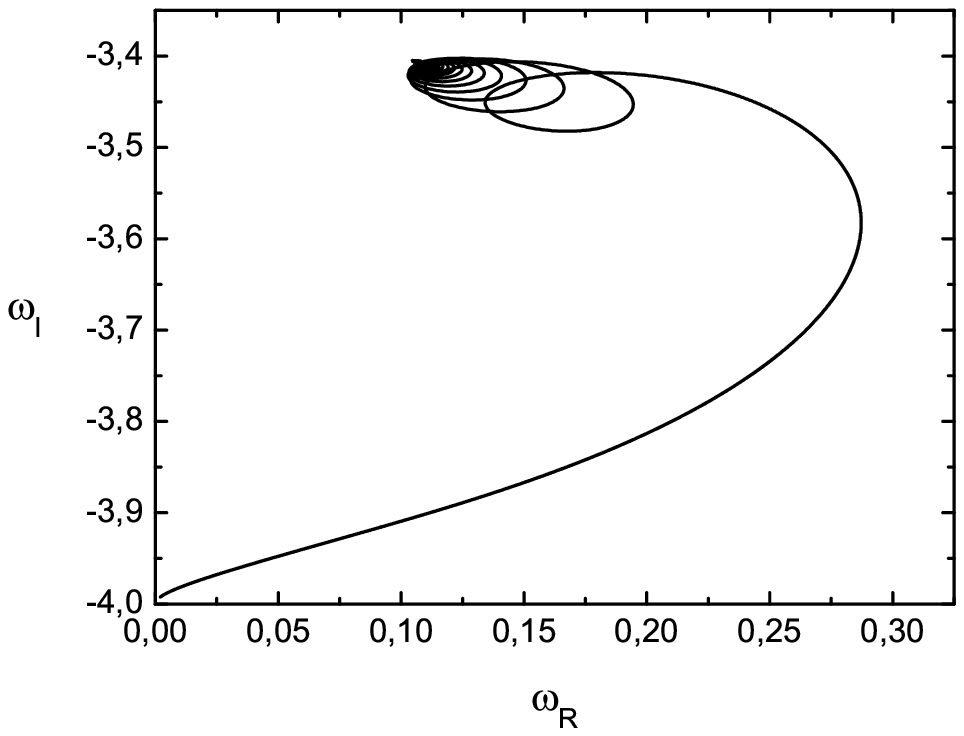,width=8cm,angle=0} \\
\end{tabular}
\caption{The left panel shows the trajectories described in the
complex-$\omega$ plane by the doublets emerging close to the
Schwarzschild algebraically special frequency ($\tilde \Omega_2=-4i$)
when $m>0$ and $l=2$. Notice that the real part of modes with $m>0$
tends to $\omega_R=m$ as $a\to 1/2$.  The right panel shows the
spiralling trajectory of the mode with $m=0$.
\label{fig:fig8}}
\end{center}
\end{figure*}

Maassen van den Brink suggested (see note [46] in \cite{MVDBas}) that
quasinormal modes corresponding to the algebraically special frequency
with $m>0$ may have one of the following three behaviors in the
Schwarzschild limit: they may merge with those having $m<0$ at a
frequency $\tilde \Omega'_l$ such that $|\tilde \Omega'_l|<|\tilde
\Omega_l|$ (but $|\tilde \Omega'_l|>|\tilde \Omega_l|$ for $l\geq 4$)
and disappear, as suggested by Onozawa \cite{O}; they may terminate at
some (finite) small $a$; or, finally, they may disappear towards
$\omega=-\ii\infty$.  Recently Maassen van den Brink {\it et al.}
\cite{MVDBdoublet} put forward another alternative: studying the
branch cut on the imaginary axis, they found that in the Schwarzschild
case a pair of ``unconventional damped modes'' should exist. For $l=2$
these modes were identified by a fitting procedure to be located on
the unphysical sheet lying behind the branch cut (hence the name
``unconventional'') at
\be\label{unconv}
\omega_\pm=\mp0.027+(0.0033-4)\ii.  
\ee 
An approximate analytical calculation confirmed the presence of these
modes, yielding
\be 
\omega_+\simeq-0.03248+(0.003436-4)\ii, 
\ee 
in reasonable agreement with (\ref{unconv}).  If their prediction is
true, an {\it additional} quasinormal mode multiplet should emerge
near $\tilde \Omega_l$ as $a$ increases. This multiplet {\it ``may
well be due to $\omega_\pm$ splitting (since spherical symmetry is
broken) and moving through the negative imaginary axis as $a$ is
tuned''} \cite{MVDBdoublet}.  In the following paragraph we will show
that a careful numerical search indeed reveals the emergence of such
multiplets, but these do not seem to behave exactly as predicted in
\cite{MVDBdoublet}.

\begin{figure*}
\begin{center}
\begin{tabular}{cc}
\epsfig{file=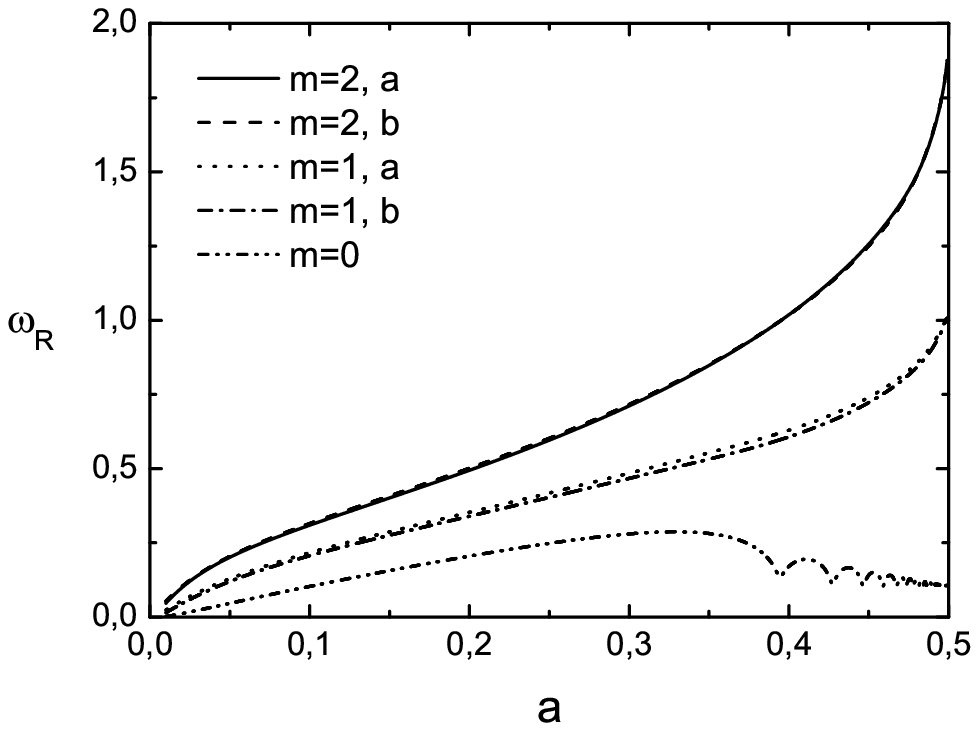,width=8cm,angle=0} &
\epsfig{file=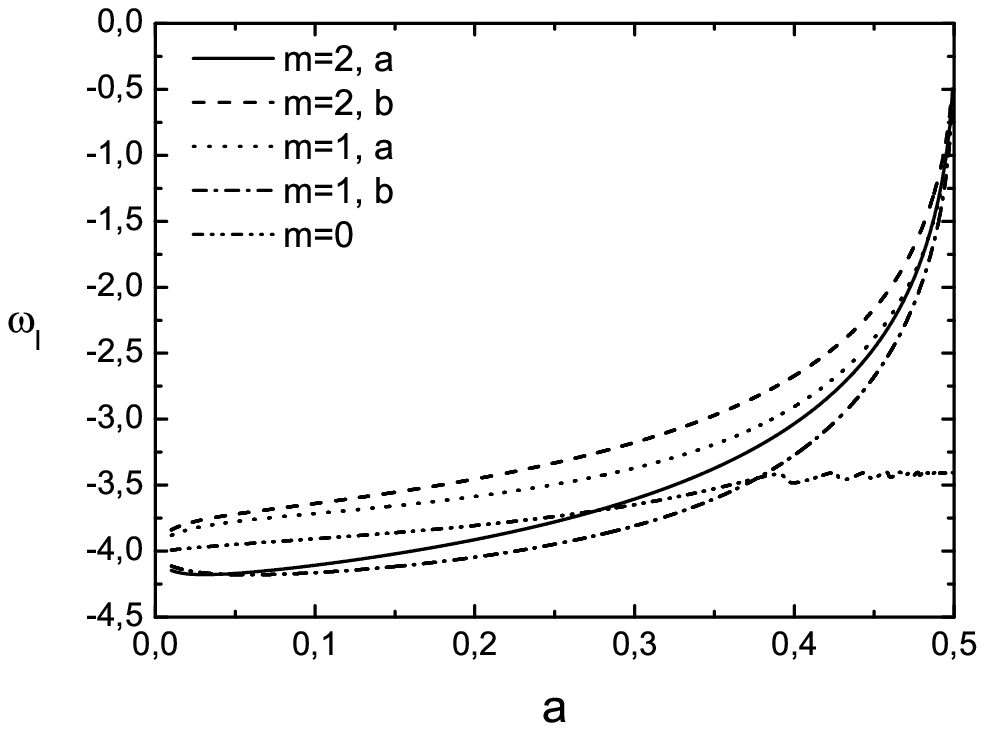,width=8cm,angle=0} \\
\epsfig{file=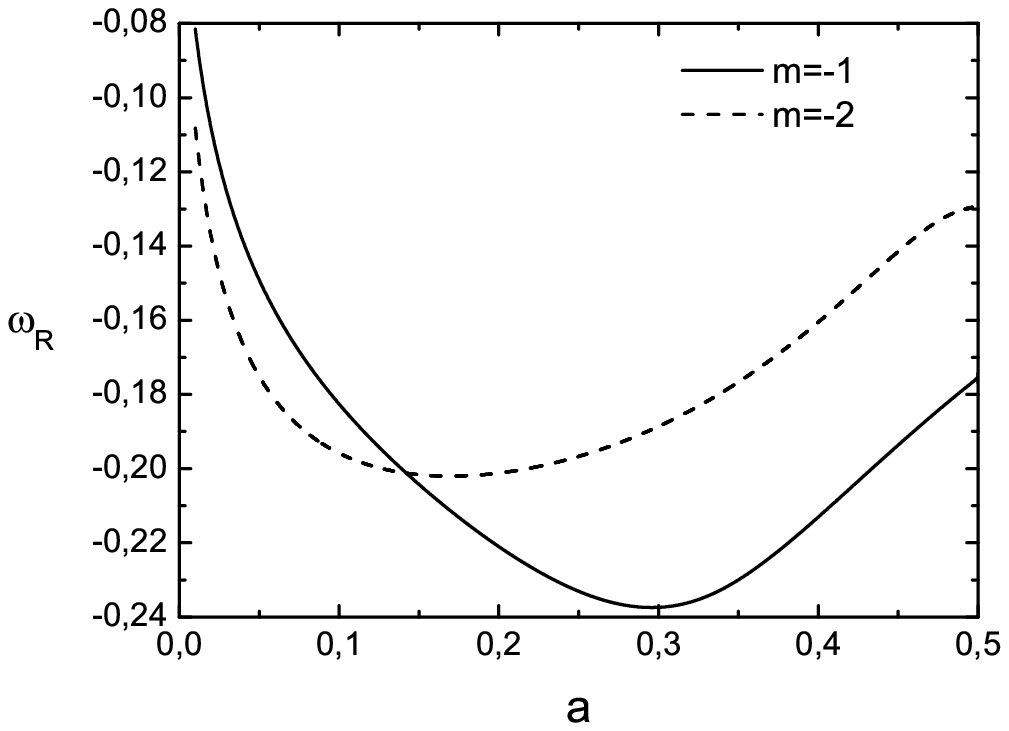,width=8cm,angle=0} &
\epsfig{file=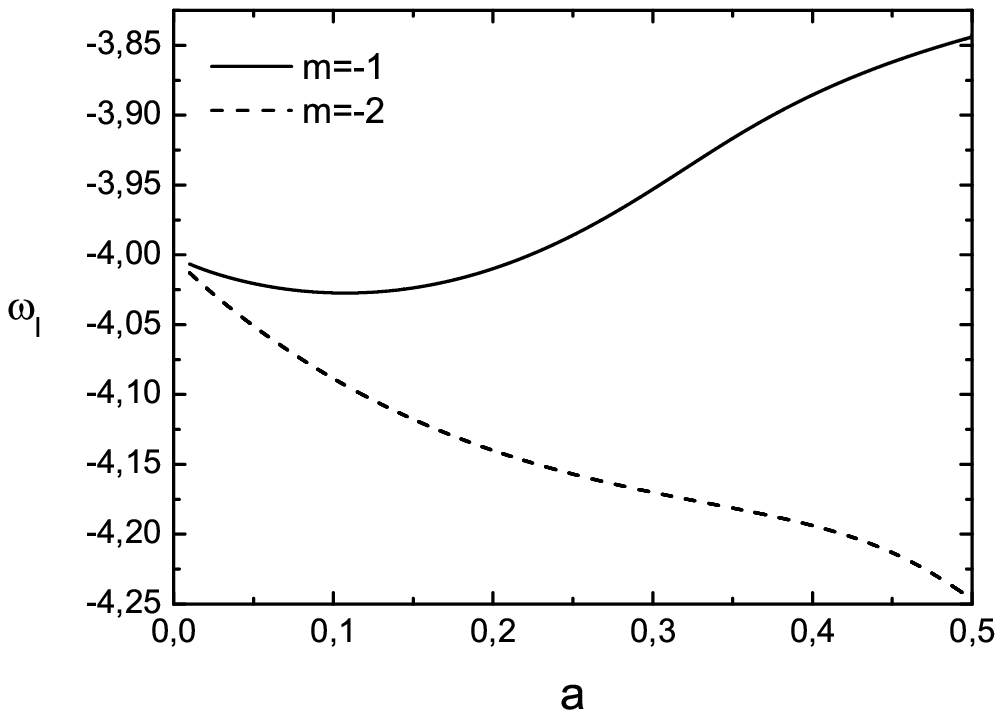,width=8cm,angle=0}
\end{tabular}
\caption{The top row shows the real and imaginary parts (left and
right, respectively) of the ``doublet'' of quasinormal modes emerging
from the algebraically special frequency as functions of $a$.  The
doublets only appear when $m>0$. We also overplot the real and
imaginary parts of the mode with $l=2$, $m=0$ (showing the usual
oscillatory behavior). The bottom row shows, for completeness, the
real and imaginary parts (left and right, respectively) of modes with
negative $m$ branching from the algebraically special frequency.
\label{fig:fig9}}
\end{center}
\end{figure*}

\subsection{Numerical search and quasinormal mode multiplets}

As we have seen from the summary in the previous paragraph, the
situation for Kerr modes branching from the algebraically special
Schwarzschild mode is not clear. There are still many open
questions. Is a multiplet of modes emerging from the algebraically
special frequency when $a>0$? Can quasinormal modes be matched by the
analytical prediction (\ref{VDBsmalla}) at small values of $a$?  If a
doublet does indeed appear, as recently suggested in
\cite{MVDBdoublet}, does it tend to the algebraically special
frequency $\tilde \Omega_2=-4\ii$ as $a\to 0$, does it tend to the
values predicted by formula (\ref{unconv}), or does it tend to some
other limit?

After carrying out an extensive numerical search, we have found some
surprises. Our main result is shown in the left panel of Figure
\ref{fig:fig8}. There we show the trajectories in the complex plane of
quasinormal modes with $l=2$ and $m>0$: a {\it doublet} of modes does
indeed appear close to the algebraically special frequency. Both modes
in the doublet tend to the usual limit ($\tilde \Omega_2=m$) as $a\to
1/2$. We have tried to match these ``twin'' modes with the predictions
of the analytical formula (\ref{VDBsmalla}). Unfortunately, none of
the two branches we find seems to agree with (\ref{VDBsmalla}) at
small $a$.  We could only find a mode doublet for $m>0$. For $m\leq 0$
the behavior of the modes is, in a way, more conventional. For
example, in the right panel of Figure \ref{fig:fig8} we see the $l=2$,
$m=0$ mode emerging from the standard algebraically special frequency
$\tilde \Omega_2$ and finally describing the ``usual'' spirals as $a$
increases.

In the top left panel of Figure \ref{fig:fig9} we see that the real
part of the twin modes with $m\geq 0$ goes to zero as $a\to 0$ with an
$m$--dependent slope. However, the top right panel in the same Figure
shows that the imaginary part of the modes does {\it not} tend to $-4$
as $a\to 0$. Qualitatively this behavior agrees rather well with that
predicted by equation (\ref{unconv}).  Extrapolating our numerical
data to the limit $a\to 0$ yields, however, slightly different
numbers. Our extrapolated values for $l=2$ are $\omega=(-4-0.10)\ii$
and $\omega=(-4+0.09)\ii$.

At present, we don't know why the doublet only appears when
$m>0$. This fact is confirmed by numerical searches close to the
algebraically special frequency $\tilde \Omega_3$ for $l=3$. Once
again, a quasinormal mode multiplet only appears when $m>0$. In
particular, we see the appearance of a doublet similar to the modes
shown in the left panel of Figure \ref{fig:fig8}. Extrapolating
numerical data for the $l=3$ doublet in the limit $a\to 0$ yields
$\omega=(-20-0.19)\ii$ and $\omega=(-20+0.24)\ii$.

A more careful search near the algebraically special frequency $\tilde
\Omega_3$ surprisingly revealed the existence of other quasinormal
modes. However, the additional modes we find may well be ``spurious''
modes due to numerical inaccuracies, since we are using the continued
fraction technique in a regime ($|\omega_I|\gg 1$, $\omega_R\simeq 0$)
where it is not supposed to work well.

\section{Conclusions and outlook}
\label{sec:conc}

The quasinormal mode spectrum of Schwarzschild, RN and Kerr black
holes is overwhelmingly rich and complex. Our understanding has
certainly improved over time, but many mysteries are still unsolved.
%The Cheshire cat is as elusive as ever! 
The following is a (partial and necessarily biased) list of open
problems.

\begin{itemize}

\item[1)]
A general lesson from the study of charged and rotating black holes
seems to be: the spectrum of the limit is not the limit of the
spectrum. This is a well-known fact in spectral theory, but it is
quite striking in the black hole context. ``Ordinary'' black hole
quasinormal modes suggest that there should be some continuity between
the Schwarzschild and Kerr (RN) solutions as $a\to 0$ ($Q\to 0$). At a
closer look, the asymptotic spectrum in the limit $|\omega_I|\to
\infty$ seems to violate this continuity requirement.  What is the
mathematical and (more importantly) physical motivation of this
discontinuity?

\item[2)]
How does the picture change when we include {\it both} rotation and
charge? As anticipated in the introduction, the main difficulty here
is that perturbation equations for Kerr-Newman black holes are
non-separable (unless we introduce some {\it ad hoc}
approximations). Work in this direction is ongoing \cite{knewman}.

\item[3)]
The algebraically special mode is a mystery on its own. Numerical
methods show the presence of a quasinormal mode {\it close to} (but
not quite {\it at}) the algebraically special frequencies determined
by Chandrasekhar -- this is true both in the Schwarzschild and in the
extremal RN cases (see Figures \ref{fig:fig1} and
\ref{fig:figxrn}). In the Kerr case, a {\it doublet} of modes with
$m>0$ comes out of the algebraically special Schwarzschild frequency,
but no mode splitting is observed when $m\leq 0$. Analytical work on
the ``supersymmetry breaking'' occurring at the algebraically special
frequency only partially clarifies the situation \cite{MVDBas}. We
definitely need a better understanding of the meaning of quasinormal
modes and total transmission modes in this situation.

\item[4)]
The behavior of Kerr quasinormal frequencies with $m>0$ is now better
understood. Not all modes tend to the critical frequency for
superradiance in the limit $a\to 1/2$. Indeed, for some modes
$\omega_I$ does not even tend to zero in the extremal limit (see
Figures \ref{fig:fig4ono} and \ref{fig:fig13}). The explanation seems
to be that some implicit assumptions in Detweiler's original treatment
were probably overlooked \cite{De,vitor}.

\item[5)]
Our present physical understanding of highly-damped black hole
oscillations is admittedly very poor. What is the meaning of the
spirals we see for RN quasinormal modes (and for Kerr modes with
$m=0$)? Is the ``universal'' Kerr asymptotic frequency $\varpi(a)$
(Figure \ref{fig:fig13}) somehow connected to other fundamental
frequencies of the Kerr spacetime, eg. last stable photon orbits? Why
does the RN asymptotic spectrum depend on the causally disconnected
region inside the horizon? Is the same true of the Kerr spacetime? And
what about Kerr-Newman black holes?
\end{itemize}

At a more fundamental level, we should go back to the original
motivations behind this investigation. The complexity of the (purely
classical) black hole oscillation spectrum is striking. Can we still
believe that there is a link between asymptotic quasinormal
frequencies and black hole quantization, or was it just a numerical
coincidence? Can we really think of black holes as ``quantum gravity
atoms'' \cite{hod,bek}, being the simplest objects we can build out of
gravity alone? If the analogy between classical black hole
oscillations and atomic spectra holds, the bizarre nature of the
quasinormal mode spectrum is telling us that there's a long way to go
before we fully understand how to quantize gravity.

% \vskip 12pt
%
% {\it ``Please would you tell me,'' said Alice, a little timidly, for
% she was not quite sure whether it was good manners for her to speak
% first, ``why your cat grins like that?''
% 
% ``It's a Cheshire cat,'' said the Duchess, ``and that's why.''
% 
% [Alice] took courage, and went on again:--
% ``I didn't know that Cheshire cats always grinned; in fact, I didn't
% know that cats COULD grin.''
% 
% ``They all can,'' said the Duchess; ``and most of 'em do.''
% 
% ``I don't know of any that do,'' Alice said very politely, feeling
% quite pleased to have got into a conversation.
% 
% ``You don't know much,'' said the Duchess; ``and that's a fact.''}
%
% \vskip 12 pt

\section*{Acknowledgements}

This short review is based on a talk given at the ``Workshop on
Dynamics and Thermodynamics of Black Holes and Naked Singularities''
(Milan, May 2004) and on work done jointly with Vitor Cardoso, Kostas
Kokkotas, Hisashi Onozawa and Shijun Yoshida. I am especially grateful
to Vitor and Kostas for many comments and corrections. This work was
supported in part by the National Science Foundation under grant PHY
03-53180.


\begin{thebibliography}{00}
\bibitem{N} H.-P. Nollert, Phys. Rev. D {\bf 47}, 5253 (1993). Notice that this
paper uses the opposite convention on the Fourier transform with
respect to Leaver's \cite{L}, so that stable modes have positive
imaginary part $\omega_I$. One should replace $\omega\to -\omega$ in
Nollert's equations for consistency with Leaver's conventions (that we
follow here).
\bibitem{modeorder} Our numerical codes are in excellent agreement
with the results in \cite{AO}. When comparing with their paper note,
however, that Andersson and Onozawa count modes starting from $n=0$,
while we label the fundamental mode by $n=1$, following Leaver
\cite{L}.
\bibitem{unitsMN} For computational convenience, Motl and Neitzke
\cite{MN} fixed units in a somewhat unconventional way. They
introduced a parameter $k$ related to the black hole charge and mass
by $Q/M=2\sqrt{k}/(1+k)$, so that $\beta=4\pi/(1-k)=1/T_H$ is the
inverse black hole Hawking temperature and $\beta_I=-k^2\beta$ is the
inverse Hawking temperature of the inner horizon.
\bibitem{comment} In \cite{breuerbook} an attempt was made to
generalize Flammer's large-$(a\omega)$ expansion for general spin
$s$. However, the result given there for the number of zeroes of the
angular wavefunction (page 115) is wrong: see \cite{casals} and
Theorem 4.1 in \cite{breuer}. Therefore, although the derivation is
essentially correct, the expression for the leading term given in
\cite{breuerbook} is not.
\end{thebibliography}
\end{document}